\documentstyle[a4,12pt,axodraw]{article}

\def\pd{\partial}
\def\a{\alpha}
\def\b{\beta}
\def\dl{\delta}
\def\s{\sigma}

\def\eps{\epsilon}
\def\lam{\lambda}
\def\Lam{\Lambda}
\def\bg{{\bar g}}
\def\hg{{\hat g}}

\def\tt{{\tilde t}}
\def\te{{\tilde e}}
\def\tb{{\tilde b}}
\def\bnabla{{\bar \nabla}}
\def\hnabla{{\hat \nabla}}
\def\bR{{\bar R}}
\def\hR{{\hat R}}
\def\bF{{\bar F}}
\def\bG{{\bar G}}
\def\bE{{\bar E}}
\def\bDelta{{\bar \Delta}}
\def\bBox{\stackrel{-}{\Box}}
\def\bpsi{{\bar \psi}}

\def\gm{\gamma}
\def\Gm{\Gamma}
\def\om{\omega}
\def\sq{\sqrt}
\def\e{\hbox{\large \it e}}
\def\half{\frac{1}{2}}
\def\fr{\frac}
\def\pp{\prime}
\def\arr{\rightarrow}
\def\bb{\begin{equation}}
\def\ee{\end{equation}}
\def\bba{\begin{eqnarray}}
\def\eea{\end{eqnarray}}

\begin{document}

\begin{titlepage}

\begin{tabbing}
   qqqqqqqqqqqqqqqqqqqqqqqqqqqqqqqqqqqqqqqqqqqqqq 
   \= qqqqqqqqqqqqq  \kill 
         \>  {\sc KEK-TH-806 }    \\
         \>       hep-th/0203250 \\
         \>  {\sc March, 2002} 
\end{tabbing}

\vspace{1.5cm}

\begin{center}
{\Large {\bf Resummation and Higher Order Renormalization in   
4D Quantum Gravity}}
\end{center}

\vspace{1.5cm}

\centering{\sc Ken-ji Hamada\footnote{E-mail address : 
hamada@post.kek.jp} }

\vspace{1cm}

\begin{center}
{\it Institute of Particle and Nuclear Studies, \break 
High Energy Accelerator Research Organization (KEK),} \\ 
{\it Tsukuba, Ibaraki 305-0801, Japan}
\end{center} 

\vspace{7mm}

\begin{abstract} 
Higher order renormalization in 4D quantum gravity 
is carried out using dimensional regularization with 
great care concerning the conformal-mode dependence. 
In this regularization, resummation can be automatically 
carried out without making an assumption  
like that of David, Distler and Kawai. 
In this paper we consider a model of 4D quantum gravity 
coupled to QED. Resummation inevitably implies a four-derivative 
quantum gravity. 
The renormalizability is directly checked up to $O(e_r^6)$ 
and $O(t_r^2)$, where $e_r$ and $t_r$ are the running coupling 
constants of QED and the traceless gravitational mode. 
There is no other running coupling constant in our model. 
The conformal mode is treated exactly, which means it is unrenormalized.
It is found that Hathrell's results are included in our results.  
As a by-product, it is found that a higher-order gravitational 
correction to the beta function of QED is negative.  
An advantage of our model is that in the very 
high-energy regime,  it closely resembles exactly solvable 2D 
quantum gravity. Thus, we can study physical states of 
4D quantum gravity in this regime in parallel to those 
of 2D quantum gravity, which can be described with 
diffeomorphism invariant composite fields.  


\end{abstract}
\end{titlepage}  

\section{Introduction and Summary}
\setcounter{equation}{0}
\noindent

Einstein's general relativity has been extremely well tested, 
and it has been believed that the graviton exists in a classical 
description and to obtain a quantum description, it 
should be quantized~\cite{d,tv,w}.  
However, the existence of the graviton at the quantum level  
has not yet been conclusively demonstrated. 
It is well known that Einstein's theory is 
unrenormalizable and that the action is unbounded below.
Also, it is doubtful whether the usual graviton picture 
can be applied beyond the Planck mass scale, where the ordinary 
particle picture seems to become invalid gravitationally. 

In an attempt to resolve the problem of renormalizability, 
a four-derivative gravity~\cite{ud}--\cite{h01} 
has been investigated, because in this case the gravitational 
coupling constant becomes dimensionless, 
and we can avoid the unboundedness problem of 
the gravitational action. Furthermore, in four-derivative gravity, 
neither infrared (IR) effects nor ultraviolet 
(UV) effects can be ignored.  
This is in accordance with the concept of background-metric independence, 
because it implies a loss of physical distance. 
For these reasons, four-derivative gravity is prefered as 
a quantum theory of gravity that is independent of the background 
metric~\cite{am}--\cite{h01}.  

However, many problems remain to be solved in existing  
four-derivative gravity models. One is renormalizability. 
Although these models are renormalizable in the sense of  
power-counting~\cite{dp}--\cite{eor}, 
this does not imply renormalizability in the true sense. In fact, 
it has been pointed out that inconsistencies will 
appear at higher orders~\cite{r,ft84}.
The second problem is that the $R^2$ action introduced to make the action 
bounded below is not asymptotically free~\cite{ft82}.
The third problem is unitarity~\cite{t,ft82,bhs}.  
There is an idea regarding unitarity proposed by Tomboulis~\cite{t} 
based on the argument of Lee and Wick~\cite{lw}. 
In short, this idea is that gravitational quantum corrections move 
a ghost pole to a pair of complex poles on the physical sheet. 
In relativistic theory, it is known that a vertex decaying from 
real-pole states to such complex-pole states gives 
measure-zero contributions. 
Thus, there is no vertex decaying 
into ghosts in 4D quantum gravity.  The proof of unitarity seems, 
however, nonrigorous.  
To begin with, it is not understood what constitutes a physical state 
in 4D quantum gravity,   
as the strong IR behavior seems to make physical states change.  
Ghosts might be concealed by the physical state 
condition, reflecting the background-metric independence.  
In this paper we investigate these issues, especially the problems of 
renormalizability and the nature of physical states. 

Recently, through studies of 
2D quantum gravity~\cite{p}--\cite{kkn}, 
it has come to be understood that all of the problems mentioned 
above arise because 
the formulation does not correctly preserve diffeomorphism invariance. 
In this paper, we show that dimensional 
regularization~\cite{tv}, carried out  
{\it with great care concerning the conformal-mode 
dependence}~\cite{kkn,hath-scalar,hath-QED,f},  
is advantageous because it manifestly preserves diffeomorphism invariance 
and that the resummation process explained 
in the next section can be 
automatically carried out without making an assumption like that made in  
the procedure of David, Distler and Kawai (DDK)~\cite{dk}. 

Near two dimensions, the $D$-dimensional action is almost unique, 
and it is given by the scalar curvature, $R$.  
However, near four dimensions, there is an indefiniteness in 
$D$-dimensional gravitational actions.  
To determine the actions here, we need extra 
conditions other than diffeomorphism invariance.
Our proposal is that there are $D$-dimensional actions, 
$F_D$ and $G_D$, which become, respectively, {\it relevant} and 
{\it marginal} at the quantum level, 
and by using these actions we can construct a renormalizable 
model described by 
$$
    I=\int d^D x \sq{g} \biggl\{
        \fr{1}{t^2} F_D + b G_D +\fr{1}{4}F_{\mu\nu}F^{\mu\nu}
        + \sum_{j=1}^{n_F} i{\bar \psi}_jD\!\!\!\!/ \psi_j 
        -m^2 R + \Lam  \biggr\}     
$$
[which is (\ref{bare-action}), below]. 
The explicit forms of $F_D$ and $G_D$ 
are presented in Section 3. 
The $F_D$ term is naturally given by the square 
of the $D$-dimensional Weyl tensor, while the  marginal 
term, $G_D$, which is a combination reduced to the Euler  
density at $D=4$, is rather non-trivial.  
To determine the form of $G_D$,   
we consider the resemblance of the 2D quantum gravity 
action, $R$, near two dimensions and $G_D$ near four dimensions. 
Furthermore, considering the integrability condition~\cite{bcr} 
for conformal anomalies~\cite{cd}--\cite{ds} 
near four dimensions, we are able to determine $G_D$. 

All bare coupling constants and fields, except for the bare 
constant, $b$, and the conformal mode, are renormalized 
in the usual manner by introducing a renormalization factor,
while the conformal mode, which is treated 
exactly, is not renormalized~\cite{h00} 
and $b$, which is composed of purely pole terms, 
is treated  in a slightly different way. 
This model has two coupling constants: the QED interaction, $e$, 
and  the self-interaction of the traceless 
gravitational mode, $t$. 
The running coupling constants are only their renormalized 
versions, $e_r$ and $t_r$.  
The bare constant $b$ is not an independent coupling constant. 
It is expanded in the coupling constants $t$ and $e$, 
and the effective action associated with 
the $bG_D$ term  has a {\it scale-invariant} form~\cite{r}.  
These points are discussed in Sections 4 and 5. 
The renormalizability is directly 
checked up to $O(e_r^6)$ and $O(t_r^2)$ in Section 6. 
We find that Hathrell's results~\cite{hath-QED} 
are contained in our results. 

The beta function for the renormalized coupling constant, $t_r$, 
is given by 
$$
      \beta_t = - \biggl(\fr{n_F}{40}+\fr{10}{3} \biggr)
                   \fr{t_r^3}{(4\pi)^2} 
                -\fr{7 n_F}{288}\fr{e_r^2 t_r^3}{(4\pi)^4} 
                +O(t_r^5)   
$$ 
[which is (\ref{beta-t})]. 
Thus, the theory is asymptotically free with respect to $t_r$. 
For the renormalized coupling constant, $e_r$, we obtain the 
beta function     
$$
    \b_e = \fr{4 n_F}{3}\fr{e_r^3}{(4\pi)^2} 
           + \biggl( 4 n_F - \fr{8}{9}\fr{n_F^2}{b_c}
                \biggr) \fr{e_r^5}{(4\pi)^4} 
           + O(e_r^3t_r^2, e_r t_r^4)  
$$
[which is (\ref{beta-e})]. Here, the term proportional to $1/b_c$  
is a gravitational correction, and $b_c = 11 n_F/360 + 40/9$. 
Note that the $O(e_r^5)$ term becomes negative for $n_F \geq 24$. 
The results for $\b_t$ and $\b_e$ seem to suggest that 
all terms in the beta functions 
related to gravity are negative.  
Hence, provided that the coefficients of $O(e_r^3 t_r^2)$ 
and $O(e_r t_r^4)$ in $\b_e$ are negative, 
we can make $e_r$ and $t_r$ sufficiently small asymptotically 
at very high energies,   
when the running starts from near the Planck mass scale, 
where both coupling constants (especially $t_r$) may be large.      

An advantage of this model is that we can exactly prove 
background-metric independence for the conformal mode 
in the limit that all running coupling 
constants vanish~\cite{amm1,hs}. 
Such a limit may be realized in the very high-energy region.   
This free field model is completely analogous to 2D quantum gravity. 
As is well known, 2D quantum gravity can be solved exactly~\cite{kpz,dk},  
and its physical states can be classified completely using operator 
formalism~\cite{bmp}.   
In four dimensions, it is not simple to construct an operator 
formalism in which composite fields can be treated. 
In dimensional regularization, however, there is a  
a close relation between them, and for this reason, physical 
states in 4D quantum gravity can be investigated in analogy to 
those in 2D quantum gravity, as discussed in Section 7. 
The last section is devoted to conclusions and discussion.

\section{Resummation in Quantum Gravity}
\setcounter{equation}{0}
\noindent

In this section we summarize the idea of resummation in 4D quantum gravity. 
In general, the idea can be applied to any even-dimensional quantum gravity 
system. 

We first give a general argument regarding resummation in {\it exactly} four 
dimensions.  
Diffeomorphism invariant quantum gravity is formally defined by  
functional integration over the metric, 
\bb
   Z= \int \frac{[g^{-1}dg]_g [df]_g}{\hbox{vol(diff.)}} 
          \exp \{-I(f,g)\}, 
              \label{z1}
\ee
where $f$ is a matter field and $I$ is an invariant action. The measure of 
the metric is defined by the norm
\bb
    \| g^{-1}d g \|^2_g  = \int d^4 x \sq{g} tr (g^{-1}dg)^2, 
              \label{measure}              
\ee
where $(g^{-1}dg)^{\mu}_{~\nu}= g^{\mu\lam}d g_{\lam\nu}$. 
  
To formulate a well-defined quantum theory of gravity, we must introduce 
a non-dynamical background metric, $\hg$, and use the measure defined 
on this background metric, which is obtained by replacing $\sq{g}$ in 
(\ref{measure}) with $\sq{\hg}$. However, this replacement violates 
diffeomorphism invariance. Measures defined on $\hg$ are no longer 
invariant under a general coordinate transformation. 
To recover diffeomorphism invariance, we must add a {\it local} action, 
$S$. Hence, we have   
\bb
   Z= \int \frac{[g^{-1}dg]_{\hg} [df]_{\hg}}{\hbox{vol(diff.)}} 
          \exp (-S-I). 
              \label{z2}
\ee
This is the idea of DDK~\cite{dk}, 
which was first developed in two dimensions~\cite{s,h93,bmp} 
and then generalized to four dimensions~\cite{am}--\cite{h01}.   
The local action $S$ is a four-derivative quantum mechanical quantity 
that can be considered a contribution induced from the measures. 
When we quantize gravity, we must quantize this local action again   
as a tree action and develop a perturbation theory. 
We here call this procedure {\it resummation in quantum gravity}. 

Resummation works well in two dimensions, 
where the traceless mode is non-dynamical, so that 
2D quantum gravity can be described as a free theory. 
The idea of DDK can be put into practice using a procedure 
in which, after assuming the form of the local action,  
its overall coefficient is determined from the condition 
that the theory becomes background-metric independent.    
However, in four dimensions, the traceless mode becomes  
dynamical, and there is no method to treat it, except for 
perturbation.   
The DDK method becomes unclear at higher orders of 
the perturbation. Thus, we need a more rigorous method to
perform resummation without making any assumption.
 
The following important observation is believed to be true:  
When we use dimensinal regularization, the results are independent 
of how the measure is chosen, because in this regularization 
$\delta^D (0)=\int d^D p =0$. 
Here, a question arises: Where has the local action $S$ gone? 
In this paper we show that information regarding $S$ is completely 
included between $D$ and $4$ dimensions.   
Thus, dimensional regularization carried out with great care concerning 
the conformal mode dependence  
may be a manifestly diffeomorphism invariant regularization.
This idea was first applied to 2D quantum gravity by Kawai, Kitazawa  
and Ninomiya~\cite{kkn} 
and by Hathrell~\cite{hath-scalar,hath-QED} 
to field theories in curved spacetime.

\section{Dimensional Regularization in 4D Quantum Gravity} 
\setcounter{equation}{0}
\noindent

There is a problem when we apply dimensional regularization to 
4D quantum gravity. In {\it exactly} four dimensions, we usually 
use the combinations of the curvatures
\bba
  F_4 &=& R_{\mu\nu\lam\s}R^{\mu\nu\lam\s}-2R_{\mu\nu}R^{\mu\nu}
          +\fr{1}{3}R^2, 
               \\  
  G_4 &=& R_{\mu\nu\lam\s}R^{\mu\nu\lam\s}-4R_{\mu\nu}R^{\mu\nu}+R^2, 
\eea  
and $R^2$, where $F_4$ and $G_4$ are the square of the Weyl tensor and 
the Euler density in four dimensions, respectively.   
In $D$ dimensions, however, we can generally consider arbitrary 
combinations of $R^2$, $R_{\mu\nu} R^{\mu\nu}$ and 
$R_{\mu\nu\lam\s}R^{\mu\nu\lam\s}$, which reduce to $F_4$ and $G_4$ 
at $D=4$. Here, we describe such combinations as $F_D$ and $G_D$, 
respectively.

The proposal of this paper is that 
{\it the gravitational actions $F_D$ and $G_D$ are almost uniquely 
determined such that the actions $F_D$, $G_D$ and $R^2$ become, 
respectively, relevant, marginal (or scale invariant) and irrelevant 
at the quantum level, and a renormalizable quantum 
gravity can be constructed using 
only two actions, $F_D$ and $G_D$, apart from lower derivative 
gravitational actions.}

In this paper, we consider a coupled system of gravity and QED.   
In the following, we first discuss the QED sector and then determine 
$F_D$ by noting the resemblance to the gauge field action.  
The action $G_D$ is determined by noting the analogy to 
the action of 2D quantum gravity near two dimensions and also 
using the integrability condition for conformal anomalies 
near four dimensions.  

\subsection{QED sector and $F_D$}
\noindent

We first study the QED sector in $D$ dimensions. 
Although the gauge field action is conformally invariant in exactly 
four dimensions, it is not in $D$ dimensions. 
Thus, the gauge field action becomes
\bb
    \fr{1}{4} \int d^D x \sq{g}F_{\mu\nu}F^{\mu\nu} 
    =\fr{1}{4}\int d^D x \sq{\hg}~\e^{(D-4)\phi}~\bg^{\mu\lam}\bg^{\nu\s}
          F_{\mu\nu}F_{\lam\s}.
              \label{gauge-action}
\ee
Here, $F_{\mu\nu}=\nabla_{\mu}A_{\nu}-\nabla_{\nu}A_{\mu}
= \pd_{\mu}A_{\nu}-\pd_{\mu}A_{\nu}$,
and the metric is now decomposed as~\cite{kkn} 
\bb
    g_{\mu\nu}=\e^{2\phi}\bg_{\mu\nu} 
\ee     
and 
\bb
   \bg_{\mu\nu}=(\hg \e^{th})_{\mu\nu} 
   =\hg_{\mu\lam} \biggl( \dl^{\lam}_{~\nu}+t h^{\lam}_{~\nu} 
          + \fr{t^2}{2}(h^2)^{\lam}_{~\nu} + \cdots \biggr), 
\ee   
where $h_{\mu\nu}=\hg_{\mu\lam} h^{\lam}_{~\nu}$ and $tr(h)=0$. 
The gravitational coupling constant $t$ is introduced only for 
the traceless mode.  

Because the fermion part is conformally invariant in $D$ dimensions, 
we can remove the conformal-mode dependence by properly rescaling 
the fermion fields. Hence, there is no direct coupling between 
the fermions and the conformal mode.  
 
Next, consider the action $F_D$, which gives the kinetic term of the 
traceless mode. This action is analogous to the gauge field action. 
Thus, the most natural choice of $F_D$ is that expressed in terms of  
the $D$-dimensional Weyl tensor, $C^{\lam}_{~\mu\s\nu}$, as
\bba
    F_D &=& C_{\mu\nu\lam\s}C^{\mu\nu\lam\s} 
              \nonumber  \\ 
        &=& R_{\mu\nu\lam\s}R^{\mu\nu\lam\s}
           -\fr{4}{D-2}R_{\mu\nu}R^{\mu\nu}
           +\fr{2}{(D-1)(D-2)}R^2.
\eea  
Since this satisfies the equation  
$F_D=\e^{-4\phi}{\bar F}_D$, the action becomes
\bb 
     \fr{1}{t^2} \int d^D x \sq{g} F_D 
     =\fr{1}{t^2} \int d^D x \sq{\hg}~\e^{(D-4)\phi} {\bar F}_D,   
               \label{traceless-action}
\ee 
where the action is divided by $t^2$ to make the kinetic term of the 
traceless mode independent of $t$.    
 
\subsection{Determination of $G_D$}
\noindent

Here, we first point out that $G_D$ is an action that reduces to 
the Euler  density at $D=4$. 
Hence, there is a resemblance between $G_D$ 
near four dimensions  and the scalar curvature $R$ near 
two dimensions. In order to determine $G_D$, we consider this resemblance.  
  
The scalar curvature in $D$ dimensions, which is the only possibility  
for the action of 2D quantum gravity, 
is expanded around $D=2$ as  
\bb
      \int d^D x \sq{g}R 
      = \sum^{\infty}_{n=0} \fr{(D-2)^n}{n!}S_n^{(2)} 
             \label{series(2)}
\ee
and
\bb
     S_n^{(2)}(\phi,\bg) 
     = \int d^D x \sq{\hg}~\phi^n \bigl(\bDelta_2 \phi +\bR \bigr) 
        + O(\phi^n), 
           \label{S(2)}
\ee
where $\Delta_2=-\Box=-\nabla^{\lam}\nabla_{\lam}$. 
Note that the action $S_1^{(2)}$ is the well-known Liouville action, 
or, the local part of the non-local Polyakov action~\cite{p}. 
The action for $n >1$ is a generalized form of the Liouville action. 
Its non-local form is discussed in Section 5. 

In analogy to 2D quantum gravity, here we look for a four-derivative 
diffeomorphism invariant action expressed by the following 
series around $D=4$: 
\bb
      \int d^D x \sq{g}G_D 
      = \sum^{\infty}_{n=0} \fr{(D-4)^n}{n!}S_n^{(4)}, 
             \label{series(4)}
\ee
with
\bb
     S_n^{(4)}(\phi,\bg) 
     = \int d^D x \sq{\hg}~\phi^n \bigl( 2\bDelta_4 \phi +\bE_4 \bigr) 
        + O(\phi^n). 
           \label{S(4)}
\ee
Here, the operator $\Delta_4$ and the quantity $E_4$ are  
defined by~\cite{r}  
\bb
      \Delta_4 = \Box^2 
               + 2 R^{\mu\nu}\nabla_{\mu}\nabla_{\nu} 
                -\fr{2}{3}R \Box 
                + \fr{1}{3} \nabla^{\mu}R \nabla_{\mu}   
\ee
and  
\bb
      E_4=G_4 -\fr{2}{3}\Box R.  
            \label{E4}
\ee
In exactly four dimensions, $\sq{g}\Delta_4$ becomes a conformally 
invariant operator for a scalar field, and $E_4$ satisfies the equation 
$\sq{g}E_4 =\sq{\hg}( 4\bDelta_4 \phi + \bE_4)$, which is a 
four-dimensional generalization of the equation  
$\sq{g}R =\sq{\hg}(2\bDelta_2 \phi +\bR)$ at $D=2$.     
Thus, the relation between $R$ at $D=2$ and $E_4$ at $D=4$ 
is apparent. 

The first term of the action, $S_1^{(4)}$, is the local part of 
the Riegert action~\cite{r}, 
which is a four-dimensional version of the Polyalov action.  
The non-local action for $n>1$ corresponds to a generalization 
of the Riegert action, 
which is also discussed in Section 5. 

To determine $G_D$ explicitly, here we apply the argument concerning  
the integrability condition for conformal anomalies given by 
Bonora, Cotta-Ramusino and Reina~\cite{bcr}.  
Consider the general form of the conformal variation of 
the effective action in $D$ dimensions,
\bb
    \dl_{\om}\Gm = \int d^D x \sq{g} ~\om \Bigl\{ 
        \eta_1 R_{\mu\nu\lam\s}R^{\mu\nu\lam\s}
        +\eta_2 R_{\mu\nu}R^{\mu\nu} + \eta_3 R^2 
        +\eta_4 \Box \! R  +\eta_5 F_{\mu\nu}F^{\mu\nu} \Bigr\}.  
              \label{dl-gm}
\ee
Because proper combinations of terms on the r.h.s. are just possibilities 
for counterterm actions, or bare actions, we wish 
to reduce the number of coefficients. 
To this end, we impose  
the integrability condition $[\dl_{\om_1},\dl_{\om_2}]\Gm=0$ 
in $D$ dimensions, which gives the following relation 
among the coefficients (see Appendix A): 
\bb
      4\eta_1 + D\eta_2 + 4(D-1)\eta_3 + (D-4)\eta_4 =0. 
                 \label{integrability}
\ee 
In contrast to the case for $D=4$, 
we cannot take $\eta_4$ itself to be arbitrary 
in $D$ dimensions, while $\eta_5$ remains arbitrary.  
We can see that the combination of $F_D$ and $G_4$  
satisfies equation (\ref{integrability}). 
As another independent combination, 
we here introduce a $D$-dimensional extension of what is called 
the trivial conformal anomaly, 
\bb
     M_D=\Box \!R- \fr{D-4}{4(D-1)}R^2 , 
\ee 
though it is no longer trivial in $D$ dimensions. 
This combination is obtained from the conformal 
variation of $R^2$ as  
$\dl_{\om} \int d^D x \sq{g}R^2 =-4(D-1)\int d^D x \sq{g}\om M_D$.

Here, we add a condition implied by the analogy to 
2D quantum gravity.   
Specifically, we look for a combination $E_D=G_4 +\eta M_D$ such that 
$\int d^D x \sq{g} E_D 
= \int d^D x \sq{g}( G_4-\eta\fr{D-4}{4(D-1)}R^2)$ 
possesses the property of the series (\ref{series(4)}). 
This condition determines $\eta$ uniquely 
as $\eta=-\fr{4(D-3)^2}{(D-1)(D-2)}$. 
(The details of the calculation are presented in Appendix B.)     
Hence, we obtain   
\bb
     E_D = G_D -\fr{4(D-3)^2}{(D-1)(D-2)} \Box\!R ,
\ee
where 
\bb
    G_D =G_4 +\fr{(D-3)^2(D-4)}{(D-1)^2(D-2)}R^2. 
               \label{GD-def}
\ee 
Here, $E_D$ is a $D$-dimensional generalization of 
$E_4$ (\ref{E4}), and $G_D$ is just what we seek. 
As in the cases of $F_D$, $G_4$ and $F_{\mu\nu}F^{\mu\nu}$, 
the conformal variation of the spacetime integral of $E_D$ 
is proportional to $E_D$ itself multiplied by $D-4$, as   
\bb
    \dl_{\om} \int d^D x \sq{g} E_D 
     = (D-4) \int d^D x \sq{g}~\om E_D. 
\ee
Thus, on the basis of the argument of the integrability condition and 
the resemblances to gauge theory and 2D quantum gravity, 
the most natural set of conformal anomalies near four dimensions 
may be $\{ F_D, E_D, M_D, F_{\mu\nu}F^{\mu\nu} \}$.  
The combinations $F_D$ and $E_D$ (or $G_D$), 
as well as conformally invariant matter 
actions,\footnote{
Mass terms are allowed, because they are harmless to UV divergences.
}
represent possibilities for the bare actions of 4D quantum gravity near four 
dimensions.  As discussed in Section 6, the higher-order computations 
of conformal anomalies in curved spacetime carried out  
by Hathrell~\cite{hath-scalar,hath-QED,f} strongly support these combinations.   

Expanding the spacetime integral of $\sq{g}G_D$ around $D=4$, 
we obtain 
\bba
  &&  \int d^D x \sq{g}G_D  
           \nonumber \\ 
  && = \int d^D x \sq{\hg} \biggl\{ 
         \bG_4 + (D-4) \biggl( 2\phi\bDelta_4 \phi +\bE_4 \phi 
                               + \fr{1}{18}\bR^2 \biggr) 
              \label{series-GD} \\ 
  && \qquad\qquad\qquad 
        +\half(D-4)^2 \biggl( 2\phi^2 \bDelta_4 \phi 
          + \bE_4 \phi^2  +O(\phi^2) \biggr) 
              \nonumber  \\ 
  && \qquad\qquad\qquad
            +O((D-4)^3) \biggr\}.  
            \nonumber            
\eea
Hence, the $O(\phi)$ term in $S_1^{(4)}$ is determined to be 
$\fr{1}{18}\bR^2$. 

We close this section by commenting  
that the actions derived above are almost unique.  
We have the freedom to change the $O(\phi^n)$ terms in $S_n^{(2)}$ 
and $S_n^{(4)}$ as $R \arr R+*(D-2)R$ near two dimensions and 
$G_D \arr G_D +*(D-4)G_D + \natural (D-4)^3 M_D$ near 
four dimensions, where $*$ and $\natural$ are arbitrary 
constants. However, as discussed in Section 7, 
the $O(\phi^n)$ terms would not affect the physical quantities, 
such as the beta functions and  anomalous dimensions. 
Hence, in this sense, they are almost unique.  
An ambiguity proportional to $F_D$ can be absorbed into the 
$F_D$ term.

\section{Renormalization}
\setcounter{equation}{0}
\noindent
 
We conjecture that the following quantum gravity model is 
renormalizable:
\bb
    I=\int d^D x \sq{g} \biggl\{
        \fr{1}{t^2} F_D + b G_D +\fr{1}{4}F_{\mu\nu}F^{\mu\nu}
        + \sum_{j=1}^{n_F} i{\bar \psi}_j D\!\!\!\!/ \psi_j 
        -m^2 R + \Lam  \biggr\}.   
               \label{bare-action}
\ee
The Dirac operator on general manifolds is defined by 
$D\!\!\!\!/=e^{\mu\a}\gm_{\a}D_{\mu}$, where $e_{\mu}^{~\a}$ is the 
vierbein field satisfying the relations 
$e_{\mu}^{~\a}e_{\nu\a}=g_{\mu\nu}$ 
and $e_{\mu\a}e^{\mu}_{~\b}=\dl_{\a\b}$, 
and the Dirac gamma-matrices on the flat manifold are 
normalized as $\{ \gm_{\a}, \gm_{\b} \}=-2\dl_{\a\b}$.  
The covariant derivative for fermions is given by 
$D_{\mu}=\pd_{\mu}+\half \om_{\mu\a\b}\Sigma^{\a\b}+ie A_{\mu}$, where 
the connection 1-form and the Lorentz generator are defined by 
$\om_{\mu\a\b}=e^{\nu}_{~\a}(\pd_{\mu}e_{\nu\b}
-\Gamma^{\lam}_{~\mu\nu}e_{\lam\b})$ 
and $\Sigma^{\a\b}=-\fr{1}{4}[\gm^{\a},\gm^{\b}]$.  
 
The renormalization is carried out in the usual way, 
apart from the $G_D$ term, by replacing 
the bare quantities in $I$ with the 
renormalized quantities multiplied by the renormalization factors.  
To make the model finite, we need the following renormalization 
factors: 
\bb
    A_{\mu} = Z_3^{1/2}A^r_{\mu}, \quad 
    \psi_j=Z_2^{1/2}\psi^r_j, \quad 
    h_{\mu\nu}=Z_h^{1/2}h^r_{\mu\nu}.
\ee   
The coupling constants of QED and the traceless mode are 
renormalized as 
\bb
    e=Z_e e_r, \quad t=Z_t t_r. 
\ee 
As discussed below, the Ward-Takahashi identity holds 
even when QED couples with quantum gravity,   
so that $Z_e=Z_3^{-1/2}$. 
The important property of this model is that the renomalization factor 
for the conformal mode is unity; 
\bb
           Z_{\phi}=1. 
             \label{Zp}
\ee
The running coupling constants of this model are only  
$e_r$ and $t_r$. 
The inverse of the gravitational constant, $m^2$, and 
the cosmological constant, $\Lam$, are also renormalized.
The renormalization of the $G_D$ term is discussed in Subsection 4.2. 

Of course, diffeomorphism invariance does not 
prohibit us from considering the extended model consisting of  
$I^{\pp}=I+\kappa \int d^D x \sq{g}R^2$, where $\kappa >0$. 
This model might be renormalized in the perturbation of $\kappa$ 
with Eq. (\ref{Zp}) preserved. At the one-loop level, it 
becomes asymptotically non-free~\cite{am}.  
On the other hand, the integrability condition seems 
to suggest that $\kappa$ has to vanish~\cite{r,ft84}. 
In any case, our assertion is that the model can be 
renormalizable at $\kappa=0$ provided that matter fields are conformally 
invariant.   

As discussed in the previous section, the action in $D$ dimensions 
has a non-trivial conformal-mode dependence apart from the fermion 
sector. As mentioned below,  
{\it the renormalization procedure is carried out by regarding,  
in the Laurent expansion of the bare action, 
the terms with non-negative powers of $D-4$ to be propagators and 
vertices and those with negative powers of $D-4$ to be counterterms.}

\subsection{New vertices in the QED sector and $F_D$}
\noindent

The renormalization factor is given by the Laurent series of poles 
whose residues are functions of $e_r$ and $t_r$. 
To begin with, $Z_3$ is taken as  
\bb
    Z_3 = 1 + \fr{x_1}{D-4}
          +\fr{x_2}{(D-4)^2} +\cdots. 
\ee
The residues are calculated in Section 6.
Using this expression, the gauge field action (\ref{gauge-action}) is 
now given by 
\bba
  &&  \fr{1}{4} \int d^D x \sq{g}F_{\mu\nu}F^{\mu\nu} 
             \nonumber  \\ 
  &&  =\fr{1}{4}Z_3\int d^D x \e^{(D-4)\phi}\bg^{\mu\lam}\bg^{\nu\s}
          F^r_{\mu\nu}F^r_{\lam\s}    
                        \\ 
  &&  =\fr{1}{4} \int d^D x \biggl\{ 
          \biggl(1+ \fr{x_1}{D-4}+\fr{x_2}{(D-4)^2}+\cdots \biggr) 
              F^r_{\mu\nu}F^r_{\mu\nu}  
                \nonumber  \\ 
  && \qquad\qquad    
          +\biggl(D-4 + x_1+\fr{x_2}{D-4}+\cdots \biggr) 
             \phi F^r_{\mu\nu}F^r_{\mu\nu}  
                \nonumber  \\ 
  && \qquad\qquad    
          +\fr{1}{2}\biggl( (D-4)^2 + (D-4)x_1+ x_2+\cdots \biggr) 
             \phi^2 F^r_{\mu\nu}F^r_{\mu\nu} 
                \nonumber \\ 
  && \qquad\qquad 
             + \cdots \biggr\},    
                \nonumber
\eea
where we take $\hg_{\mu\nu}=\dl_{\mu\nu}$, and the same lower 
spacetime indices on the r.h.s. denotes contraction 
by $\dl_{\mu\nu}$. In the above expression, 
we omit the interaction terms between the gauge 
field and the traceless mode, which can be easily derived from the 
definition. 

Similarly, the action of the traceless mode, $\sq{g}F_D$, 
can be expanded in terms of the renormalized quantities 
using expression (\ref{traceless-action}) 
and the renormalization factors $Z_t$ and $Z_h$.  
Although the expression becomes more complicated, the expansion is 
straightforward. We do not give it here.  

Fermion fields are conformally invariant in $D$ dimensions. 
As mentioned above, we can remove the conformal-mode dependence from 
the fermion action by replacing the fermion field, $\psi$, 
with the rescaled one, $\psi^{\pp}=\e^{\fr{(D-1)}{2}\phi}\psi$,  
so that $\int d^D x \sq{g} i\bpsi D\!\!\!\!/ \psi 
=\int d^D x \sq{\hg} i \bpsi^{\pp} {\bar D}\!\!\!\!/ \psi^{\pp}$.  
Because in dimensional regularization the result is independent of 
whether $\psi$ or $\psi^{\pp}$ is chosen, 
we quantize $\psi^{\pp}$. 
In the following, we write $\psi^{\pp}$ as $\psi$. 

The bare fermion action is expanded in the coupling $t$  
in the flat background as  
\bba
   && \int d^D x  i\bpsi {\bar D}\!\!\!\!/ \psi 
               \nonumber \\ 
   && =\int d^D x \biggl\{ i \bpsi  \gm_{\mu} \pd_{\mu}\psi 
         -i\fr{t}{4} \bigl( \bpsi \gm_{\mu}\pd_{\nu}\psi 
             -\pd_{\nu} \bpsi \gm_{\mu} \psi \bigr) h_{\mu\nu} 
                \nonumber  \\ 
   && \qquad
       +i\fr{t^2}{16} \bigl( \bpsi \gm_{\mu}\pd_{\nu}\psi 
     -\pd_{\nu} \bpsi \gm_{\mu} \psi \bigr) h_{\mu\lam}h_{\nu\lam} 
     + i\fr{t^2}{16} \bpsi \gm_{\mu\nu\lam} \psi 
            h_{\mu\s}\pd_{\lam}h_{\nu\s}
                     \\ 
   && \qquad 
      -e \bpsi \gm_{\mu} \psi A_{\mu} 
      +\fr{et}{2} \bpsi \gm_{\mu} \psi A_{\nu} h_{\mu\nu} 
      -\fr{et^2}{8} \bpsi \gm_{\mu} \psi A_{\nu} h_{\mu\lam}h_{\nu\lam} 
         \biggr\} + O(t^3), 
               \nonumber  
\eea 
where $\gm_{\mu\nu\lam}=\fr{1}{3!}(\gm_{\mu}\gm_{\nu}\gm_{\lam} 
+ \hbox{anti-sym.} )$. 
Replacing the bare quantities with the renormalized quantities, 
we obtain vertices and counterterms, as usual. 

\subsection{Kinetic term of the conformal mode and 
new vertices in $bG_D$}
\noindent

The UV divergences related to the $G_D$ term are renormalized by 
writing $b$ as 
\bb
      b=\sum^{\infty}_{n=1}\fr{b_n}{(D-4)^n}.  
              \label{b-series}
\ee
Here, we take $b_0=0$, because the tree part, $b_0 \sq{g}G_D$, 
does not have a kinetic term for the gravitational fields, 
so that it is not a coupling constant.    
Then, using expression (\ref{series-GD}), we obtain 
\bba
    && b\int d^D x \sq{g} G_D 
          \nonumber   \\
    && = \int d^D x \biggl\{ 
         \biggl( \fr{b_1}{D-4}+\fr{b_2}{(D-4)^2}
                 + \cdots \biggr) \bG_4
            \nonumber  \\ 
    && \qquad\qquad 
         + \biggl( b_1 +\fr{b_2}{D-4} +\cdots \biggr)  
          \biggl( 2 \phi \bDelta_4 \phi +\bE_4 \phi 
                      +\fr{1}{18}\bR^2 \biggr) 
                   \\            
    && \qquad\qquad
        +\half \Bigl( (D-4)b_1 +b_2 +\cdots \Bigr)   
           \Bigl( 2\phi^2 \bDelta_4 \phi +\bE_4 \phi^2 
                             +\cdots  \Bigr) 
                \nonumber \\
    && \qquad\qquad
         +\cdots \biggr\}.  
                \nonumber 
\eea

We must give some remarks concerning this counterterm. 
The first is that the kinetic term of the conformal mode is induced 
by quantum effects, whose overall coefficient is given by the 
residue of the simple pole, $b_1$. 
The residue of the double pole, $b_2$, produces 
a local counterterm for the kinetic term of the conformal mode.  
However, this does not imply that $Z_{\phi}$ differs from unity.  
The second is that, even if we take $b_0 \neq 0$, which corresponds 
to the shift $G_D \arr G_D +*(D-4)G_D$ discussed at the end 
of Subsection 3.2, the physical quantities 
are not affected by its value, as discussed in Section 7. 
The third is that $b$ is a bare quantity, though we expand 
it in the renormalized coupling constants.

\section{Higher-Order Effective Action}
\setcounter{equation}{0}
\noindent

In this section, we give brief comments on the renormalized 
higher-order effective action. 
Above, we showed that the Laurent expansions of the bare actions 
using the renormalized quantities produce new local vertices. 
The overall coefficients of these vertices are related to the 
UV divergences of higher-order diagrams. The higher-order diagrams 
also produce finite non-local terms.   
Thus, the renormalized effective action of quantum gravity, $\Gamma$, 
is given by the sum of the local renormalized vertices in $I$, 
described by $I_r$, and the non-local actions produced 
by the loop diagrams, $W$, so that $\Gamma=I_r+W$~\cite{h01}. 

For the gauge sector, the new vertex $\phi^n F^r_{\mu\nu}F^r_{\mu\nu}$  
is related to the non-local term $A_r p^2 (\log p^2/\mu^2)^n A_r$ 
in momentum space, where the spacetime indices are omitted.   
The sum of these terms gives the diffeomorphism invariant 
effective action.
Similarly, the new vertex $\phi^n \bF_D^r$  
is related to the non-local term $h_r ~p^4 (\log p^2/\mu^2)^n h_r$ 
in momentum space. 

Next, consider the effective action related to the counterterm 
of the Euler density. In two dimensions, 
the higher-order effective action related to 
the new vertices, $S_n^{(2)}$, is  
given by   
\bba
   && \int d^2 x \sq{g}R_r \overbrace{\fr{1}{\Delta^r_2}R_r 
              \fr{1}{\Delta^r_2}R_r \cdots \fr{1}{\Delta^r_2}R_r}^n
                 \nonumber \\ 
   &&= 2^n ~2 \int d^2 x \sq{\hg}~\phi^n 
           \bigl(\bDelta^r_2 \phi +\bR_r \bigr) 
                + \hbox{non-local terms}. 
                 \label{EA(2)}
\eea
The $n=1$ action is the well-known non-local 
Polyakov action~\cite{p}. 
The important property of this action is that it is scale-invariant. 
The action for $n>1$ is a natural scale-invariant generalization of 
the Polyakov action.   
Here, note that the first term on the r.h.s. is just the first term 
in the new vertices, $S_n^{(2)}$. The non-local terms can be obtained  
from higher-order diagrams. 
The $O(\phi^n)$ term in $S_n^{(2)}$ is necessary to make the theory 
diffeomorphism invariant, but it would not affect the form of 
the effective action directly. 

Now, we discuss the case of 4D quantum gravity. 
A four-dimensional version of the Polyakov action was constructed 
by Riegert~\cite{r}. 
A natural scale-invariant generalization of the Riegert action is 
\bba
   && \int d^4 x \sq{g}E_4^r \overbrace{\fr{1}{\Delta^r_4}E_4^r 
        \fr{1}{\Delta^r_4}E_4^r \cdots \fr{1}{\Delta^r_4}E_4^r}^n
                  \nonumber  \\ 
   &&= 4^n ~2\int d^4 x \sq{\hg}~\phi^n 
          \bigl(2\bDelta^r_4 \phi +\bE_4^r \bigr) 
             + \hbox{non-local terms}. 
                \label{EA(4)}
\eea  
The first term on the r.h.s. is just the first term 
in the new vertices, $S_n^{(4)}$. This is one of the reasons why we 
consider $G_D$ to satisfy Eq. (\ref{series(4)}) with (\ref{S(4)}). 
The non-local terms may be derived from higher-order loop diagrams.       
This is also investigated in Subsection 6.1 using a concrete 
expression.  

\section{Direct Check of Renormalizability} 
\setcounter{equation}{0}
\noindent

In this section, we directly check the renormalizability of the 
model proposed in the previous section up to $O(e_r^6)$ and $O(t_r^2)$.

The procedure of gauge fixing for the traceless mode is presented in 
Appendix C. Here, we employ the Feynman-type gauge.    
Then, the propagator of the traceless mode, which is described by 
a spiral line, becomes 
\bb
        \fr{D-2}{2(D-3)} \frac{1}{p^4} (I_H)_{\mu\nu, \lam\s}, 
\ee
where $I_H$ is the projection operator to the traceless mode 
(\ref{IH}). 

The propagator of the conformal mode, which is described by a solid line,  
is derived from the action $b_1 S_1^{(4)}$ as 
\bb
    \frac{(4\pi)^2}{4 b_c} \frac{1}{p^4} .  
\ee
For later use, we here introduce the coupling-independent constant  
$b_c$ by  
\bb
    b_1= \fr{1}{(4\pi)^2} \Bigl\{ b_c +O(e_r^2,~t_r^2) \Bigr\}.
             \label{bc}
\ee 
The value of $b_c$ is determined by computing one-loop diagrams.   
 
In this paper we only consider UV divergences, and do not treat 
IR divergences.  Essentially, there is no problem with  
IR divergences, because of the presence of the Einstein-Hilbert 
action and the cosmological constant, which play the role of 
diffeomorphism invariant IR regulators. Technically, we evaluate 
diagrams by replacing the fourth-order propagator $1/p^4$ with 
$1/(p^2+z^2)^2$, where $z$ is a small mass scale.  
  
In the following, we set $D=4-2\eps$. For later use, we introduce 
the dimensionless quantities described by symbols with tildes,  
such as $\tt=t \mu^{-\eps}$, $\te=e \mu^{-\eps}$,  
$\tb=b \mu^{2\eps}$, and so on, where $\mu$ is an arbitrary mass scale. 
Here, we set the dimension of the conformal mode to zero, while the 
traceless mode is of dimension $\mu^{-\eps}$. 
  
\subsection{Two- and three-point functions of the traceless mode}
\noindent

We first review the known results for the two- and three-point functions 
of the traceless mode, which are related to the counterterms of  
$\sq{\hg}\bF_D$ and $\sq{\hg}\bG_D$. 
\begin{center}
\begin{picture}(450,100)(20,-10)
\Gluon(100,50)(130,50){3}{4}
\GCirc(150,50){20}{0.5}
\Gluon(170,50)(200,50){3}{4}
\Text(150,0)[]{(a)}    

\Gluon(250,40)(280,40){3}{4}
\GCirc(300,40){20}{0.5}
  \Gluon(300,60)(300,90){3}{4}
\Gluon(320,40)(350,40){3}{4}
\Text(300,0)[]{(b)} 
\end{picture} \\ 
Fig. 1. Two- and three-point functions of the traceless mode.
\end{center} 

The renormalization factor $Z_t$ is given by  
\bb
    Z_t = 1-  \biggl(\fr{n_F}{80}+\fr{5}{3} \biggr) 
                \fr{\tt_r^2}{(4\pi)^2}\fr{1}{\eps} 
          - \fr{7 n_F}{288}\fr{\te_r^2 \tt_r^2}{(4\pi)^4}\fr{1}{\eps} 
          + O(\tt_r^4). 
                 \label{Zt}
\ee   
The $O(t_r^2)$ term is the sum of one-loop contributions from 
the QED sector and the gravitational sector. 
The contributions from the one-loop diagrams with the internal photon, 
fermion and conformal mode can be calculated from 
the two-point functions of the traceless 
mode [Fig. 1-(a)], because $Z_h$ for these diagrams is 
gauge invariant, so that the relation $Z_t=Z_h^{-1/2}$ 
is satisfied at this order.  
They give the contributions $-n_F/80-1/40$~\cite{duff} 
for the QED sector and $1/60$~\cite{amm1} for the conformal mode. 
However, the contribution to $Z_h$ from the internal 
traceless mode is, in general, gauge dependent,  
so that $Z_t \neq Z_h^{-1/2}$. 
To calculate the contributions from the traceless mode, the background 
field method~\cite{ab} is useful. 
Introducing the background traceless mode 
as $\hg_{\mu\nu}=(\e^{t{\hat h}})_{\mu\nu}$ and calculating 
the two-point functions with the external $h$ fields 
replaced by ${\hat h}$, we obtain $-199/120$~\cite{ft82} 
for $Z_t$, due to the relation 
$Z_t = Z_{{\hat h}}^{-1/2}$, where $Z_{{\hat h}}$ is the 
renormalization factor of the background field ${\hat h}_{\mu\nu}$.   
The $O(e_r^2 t_r^2)$ term is the sum of two-loop contributions from 
diagrams in which only photons and fermions propagate in the internal 
lines~\cite{dh}.
Hence, the beta function for the traceless mode coupling 
constant $t_r$ is
\bb
      \beta_t = - \biggl(\fr{n_F}{40}+\fr{10}{3} \biggr)
                     \fr{t_r^3}{(4\pi)^2} 
                -\fr{7 n_F}{72}\fr{e_r^2 t_r^3}{(4\pi)^4} 
                +O(t_r^5). 
                   \label{beta-t}
\ee 
Thus, this model is asymptotically free 
with respect to $t_r$~\cite{ft82,t}. 

  The residue of the counterterm of $\sq{\hg}\bG_D$ can be calculated 
from the three-point functions of gravitational fields. 
For the simple pole term,  it is known to be 
\bb
    \tb_1 = \fr{1}{(4\pi)^2} \biggl\{ \fr{11 n_F}{360}+\fr{40}{9}                       
          -\fr{n_F^2}{6}\fr{\te_r^4}{(4\pi)^4} +O(\tt_r^2) \biggr\}. 
\ee 
The coupling-independent term in $\tb_1$ is the sum of the contributions 
from the QED sector, $(11n_F +62)/360$~\cite{duff}, 
the conformal mode, $-7/90$~\cite{amm1}, 
and the traceless mode, $87/20$~\cite{ft82}. From this, 
the coupling-independent term $b_c$ defined by equation (\ref{bc}) is 
\bb
   \tb_c = \fr{11 n_F}{360}+\fr{40}{9}. 
              \label{bcv}
\ee
The double pole term is known to be~\cite{hath-QED}
\bb 
    \tb_2 = \fr{1}{(4\pi)^2} \biggl\{ 
      \fr{2 n_F^3}{9}\fr{\te_r^6}{(4\pi)^6}+ O(\tt_r^4) \biggr\}. 
\ee

We comment here on the diffeomorphism-invariant 
effective action. As discussed in Section 5, 
it is given by the sum of the local action $I_r$ 
and the finite contribution from loop diagrams $W$, 
which contain both local and non-local terms. 
The one-loop contribution related to  
the counterterm of $\sq{\hg}\bG_D$ has the form 
\bb
   W_G (\bg_r) =\fr{b_c}{(4\pi)^2} \int d^4 x \sq{\hg} \biggl\{ \fr{1}{8} 
    \bE_4^r \fr{1}{\bDelta^r_4} \bE_4^r -\fr{1}{18}\bR_r^2 \biggr\}. 
\ee
This scale-invariant action can be obtained from corrections 
to the three-point function of the traceless mode. 
The $\bR_r^2$ term is needed to guarantee that $W_G$ does not have any 
contributions from two-point diagrams of the traceless mode 
in the flat background. 
We can show that in the sum of the action $b_1 S_1^{(4)}$  
in $I_r$ and $W_G$, the non-invariant $\bR_r^2$ terms do 
cancel out, and the renormalized effective action becomes 
manifestly diffeomorphism invariant~\cite{h01}:
\bb
    \fr{b_c}{(4\pi)^2} S_1^{(4)}(\phi,\bg_r) + W_G(\bg_r) 
    = \fr{b_c}{8(4\pi)^2}  \int d^4 x \sq{g}  
          E_4^r \fr{1}{\Delta^r_4} E_4^r.  
\ee
This is the Riegert action, or the case of $n=1$ in (\ref{EA(4)}). 
Thus, only the first term in $S_1^{(4)}$ determines the form of the 
effective action, as mentioned in Section 5. 
 
\subsection{Two-point functions of the conformal mode}
\noindent

Because the properties of the traceless mode are analogous 
to those of gauge fields, the renormalization procedure 
for this mode can be carried out straightfowardly. 
However, diagrams related to 
the conformal mode are unusual. We study these diagrams 
in this and following subsections.  

The double-pole term of $b$ appears at $O(e_r^6)$ and, at most, $O(t_r^4)$. 
This implies that there is no counterterm for two-point functions 
of the conformal mode up to $O(e_r^4)$ and $O(t_r^2)$. 
This can be seen by direct computation.  
We first demonstrate the finiteness of the $O(t_r^2)$ correction, which 
is given by the sum of diagrams (a) and (b) in Fig. 2. 
The vertices in these diagrams are obtained from 
the action $b_1 S_1^{(4)}(\phi,\bg)$, 
and they are summarized in Appendix D. 
We can see that the UV divergences exactly cancel out~\cite{h00}. 
\begin{center}
\begin{picture}(430,100)(0,10)
\Line(100,50)(200,50)
\GlueArc(150,50)(30,0,180){3}{10.5} 
\Vertex(120,50){1}\Text(120,47)[t]{$t_r$}
\Vertex(180,50){1}\Text(180,47)[t]{$t_r$}
\Text(150,20)[]{(a)}    

\Line(250,50)(330,50)
\GlueArc(290,70)(18,-90,270){3}{15} 
\Vertex(290,50){1}\Text(290,47)[ct]{$t_r^2$}
\Text(290,20)[]{(b)} 
\end{picture} \\ 
Fig. 2. $O(t_r^2)$ diagrams contributing to $\phi \!-\! \phi$.
\end{center} 

The $O(e_r^2)$ correction is trivially finite.  
The $O(e_r^4)$ correction is given by the sum of 
diagrams (a) to (g) in Fig. 3. Here, the wavy line  
and the dashed line with arrows denote the photon and fermion, respectively.  
The subdiagram in (e), appearing as a circle with $``2"$ inside, 
denotes the sum of the two-loop diagrams for the ordinary photon 
self-energy. 
We do not here depict counterterm diagrams that subtract the 
divergences of the subdiagrams.  
In the following, such counterterm diagrams 
for subdivergences are always omitted, 
and we only depict simple-pole counterterm diagrams for direct 
subtraction, whose residues are related to the 
residues of double poles in $b$ and $Z_3$.  

{}From counting the number of loops 
and $\eps$'s at the vertices, we can easily see that 
diagrams (a) to (d) in Fig. 3 yield simple poles. 
Contrastingly, although  diagrams (e) to (g) in Fig. 3 
potentially have simple-pole divergences, these  become finite 
because of the renormalization properties of QED 
that the sum of the two-loop diagrams for the photon self-energy  
has at most a simple pole, 
and the four-point function of the photon is finite. 
Thus, we merely compute diagrams (a) to (d) in Fig. 3, and 
we can see that their sum becomes finite~\cite{hath-QED}. 
\begin{center}
\begin{picture}(380,100)(0,0)
\Line(10,50)(25,50)
\PhotonArc(45,50)(20,0,360){2}{16} 
\Line(65,50)(80,50)
\Vertex(25,50){1}\Text(22,48)[rt]{$n_F e_r^2$}
\Vertex(65,50){1}\Text(68,48)[lt]{$n_F e_r^2$}
\Text(45,10)[]{(a)}    

\Line(100,50)(116,50)
\PhotonArc(135,50)(20,120,60){2}{13} 
\DashArrowArc(135,68)(10,-90,270){1}
\Line(156,50)(170,50)
\Vertex(116,50){1}\Text(112,48)[rt]{$\eps$}
\Vertex(156,50){1}\Text(158,48)[lt]{$n_F e_r^2$}
\Text(135,10)[]{(b)}

\Line(190,50)(206,50)
\PhotonArc(230,50)(25,150,30){2}{13}
\PhotonArc(230,50)(25,69,111){2}{2.5} 
\DashArrowArc(215,67)(8,-60,300){1}
\DashArrowArc(245,67)(8,-120,240){1}
\Line(256,50)(270,50)
\Vertex(206,50){1}\Text(202,48)[rt]{$\eps$}
\Vertex(256,50){1}\Text(258,48)[lt]{$\eps$}
\Text(230,10)[]{(c)}

\Line(290,50)(305,50)
\PhotonArc(330,50)(25,-70,70){2}{8}
\PhotonArc(330,50)(25,110,250){2}{8} 
\DashArrowArc(330,73)(8,-90,270){1}
\DashArrowArc(330,27)(8,-90,270){1}
\Line(355,50)(370,50)
\Vertex(305,50){1}\Text(302,48)[rt]{$\eps$}
\Vertex(355,50){1}\Text(358,48)[lt]{$\eps$}
\Text(330,10)[]{(d)}
\end{picture} \\ 
\end{center}
\begin{center}
\begin{picture}(330,100)(0,0)
\Line(10,50)(25,50)
\PhotonArc(45,50)(20,0,360){2}{16} 
\GCirc(45,68){10}{1}
\Text(45,70)[]{2}
\Line(65,50)(80,50)
\Vertex(25,50){1}\Text(22,48)[rt]{$\eps$}
\Vertex(65,50){1}\Text(68,48)[lt]{$\eps$}
\Text(45,10)[]{(e)}

\Line(100,50)(115,50)
\Photon(115,50)(130,70){2}{4}
\Photon(115,50)(130,30){2}{4}
\DashArrowLine(130,70)(130,30){1}
\DashArrowLine(130,30)(170,30){1}
\DashArrowLine(170,30)(170,70){1}
\DashArrowLine(170,70)(130,70){1}
\Photon(170,70)(185,50){2}{4}
\Photon(170,30)(185,50){2}{4}
\Line(185,50)(200,50)
\Vertex(115,50){1}\Text(112,48)[rt]{$\eps$}
\Vertex(185,50){1}\Text(188,48)[lt]{$\eps$}
\Text(150,10)[]{(f)}

\Line(220,50)(235,50)
\Photon(235,50)(250,70){2}{4}
\Photon(235,50)(250,30){2}{4}
\DashArrowLine(250,70)(270,50){1}
        \DashLine(270,50)(290,30){1}
\DashArrowLine(290,30)(250,30){1}
\DashArrowLine(250,30)(270,50){1}
        \DashLine(270,50)(290,70){1}
\DashArrowLine(290,70)(250,70){1}
\Photon(290,70)(305,50){2}{4}
\Photon(290,30)(305,50){2}{4}
\Line(305,50)(320,50)
\Vertex(235,50){1}\Text(232,48)[rt]{$\eps$}
\Vertex(305,50){1}\Text(308,48)[lt]{$\eps$}
\Text(270,10)[]{(g)}
\end{picture} \\ 
Fig. 3. $O(e_r^4)$ diagrams contributing to $\phi \!-\! \phi$.
\end{center}

Furthermore,  using Hathrell's results~\cite{hath-QED}, 
we can demonstrate the finiteness of the $O(\e_r^6)$ two-point functions 
of the conformal mode in which only photons and fermions propagate 
in the internal lines (Fig. 4). 
For the $O(\e_r^6)$ diagrams [Fig. 4-(a)] 
not depicted explicitly, we can see that the sum of the double poles 
vanishes.  The sum of the simple poles does cancel 
with the direct counterterm of $O(e_r^6)$ [Fig. 4-(b)], which comes 
from the double-pole term of $b$, because, as shown by    
Hathrell, the divergences always appear 
with the special relation $b_2=2c_1$ at $O(e_r^6)$ if we use the set of 
counterterms $bG_4$ and $cH^2$, 
where $H=\fr{R}{D-1}$ and $c_1$ is the residue of the 
simple pole of $c$. This relation is naturally realized in our model. 
From (\ref{GD-def}), the bare action $b G_D$ implies   
$c_1=\fr{(D-3)^2}{D-2}b_2 =\half b_2 +O(D-4)$. 

\begin{center}
\begin{picture}(370,100)(0,0)
\Line(100,50)(120,50)
\CArc(135,50)(15,-90,270)
\Text(135,50)[]{$e_r^6$}
\Line(150,50)(170,50)
\Text(135,10)[]{(a)}

\Line(230,50)(270,50)
\GCirc(250,50){4}{1}
  \Line(247,53)(253,47)
  \Line(247,47)(253,53)
\Text(250,10)[]{(b)}
\end{picture} \\
Fig. 4. $O(e_r^6)$ diagrams contributing to $\phi \!-\! \phi$. 
\end{center}

The most important observation by Hathrell is that this relation 
is universal, independent of matter fields. 
Actually, it has been shown that this relation is satisfied 
in conformally interacting scalar field theory, 
by Hathrell~\cite{hath-scalar}, and also in non-Abelian gauge theory, 
by Freeman~\cite{f}. 
Our proposal is, in other words,  that this special relation 
also holds for quantized qravity.  
The fact that $Z_{\phi}$ is unity at $O(t_r^2)$ 
computed using the vertices in $b_1S_1^{(4)}$
is a consequence of the combination $G_D$.  
Thus, we see that $G_D$ is a universal combination 
in 4D quantum gravity. 

Now, there is no inconsistency between Hathrell's results and 
the renormalizablity of 4D quantum gravity. 
Although his results have often been cited as evidence for the 
unrenormalizability of 4D quantum gravity, because it has been 
believed that $R^2$ divergences must vanish  
in order for an effective action to exist~\cite{r,ft84}, 
this is not true when we use dimensional regularization.  

\subsection{On $Z_1=Z_2$}
\noindent

In this subsection, we show that new gravitational interactions 
do not violate the Ward-Takahashi identity, $Z_1=Z_2$. 

The $O(t_r^2)$ corrections to $Z_2$ by gravitational fields are 
given by two diagrams, (a) and (b) in Fig. 5,  
in which the traceless mode propagates in the 
internal line. Since fermion fields do not couple with the conformal 
mode directly, there is no diagram in which the conformal mode propagates 
in the internal line at the one-loop level.  
Hence, the sum of the two diagrams in Fig. 5 yields   
\bb
    Z_2 -1= -\fr{21}{64}\fr{\tt_r^2}{(4\pi)^2}\fr{1}{\eps} 
\ee
for all species of fermions in the Feynman-type gauge. 
\begin{center}
\begin{picture}(430,100)(0,10)
\DashArrowLine(100,50)(120,50){1}
\DashArrowLine(120,50)(180,50){1}
\DashArrowLine(180,50)(200,50){1}
\GlueArc(150,50)(30,0,180){3}{10.5} 
\Vertex(120,50){1}\Text(120,45)[t]{$t_r$}
\Vertex(180,50){1}\Text(180,45)[t]{$t_r$}
\Text(150,20)[]{(a)}    

\DashArrowLine(250,50)(290,50){1}
\DashArrowLine(290,50)(330,50){1}
\GlueArc(290,70)(18,-90,270){3}{15}
\Vertex(290,50){1}\Text(290,45)[t]{$t_r^2$}
\Text(290,20)[]{(b)} 
\end{picture} \\ 
Fig. 5. $O(t_r^2)$ corrections to $Z_2$.
\end{center} 

In contrast to the case for $Z_2$, the $O(t_r^2)$ corrections to $Z_1$ are 
given by the seven diagrams (a) to (g) in Fig. 6. 
Nevertheless, we can show that the Ward-Takahashi identity holds 
even when QED couples with quantum gravity. 
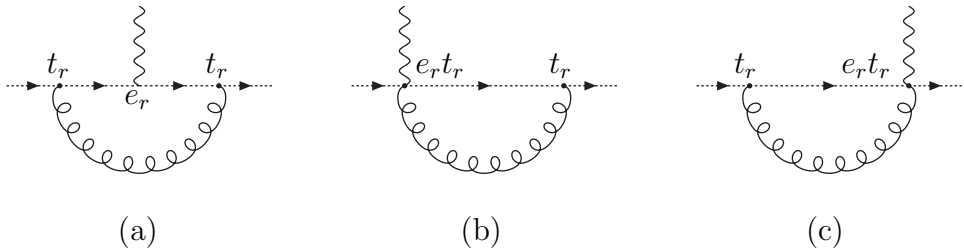
\begin{center}
\begin{picture}(510,100)(70,-15)
\DashArrowLine(100,50)(120,50){1}
\DashArrowLine(120,50)(150,50){1}
  \Photon(150,50)(150,80){2}{4}\Text(150,45)[]{$e_r$}
\DashArrowLine(150,50)(180,50){1}
\DashArrowLine(180,50)(200,50){1}
\GlueArc(150,50)(30,180,360){3}{10.5} 
\Vertex(120,50){1}\Text(120,55)[b]{$t_r$}
\Vertex(180,50){1}\Text(180,55)[b]{$t_r$}
\Text(150,-5)[]{(a)}   

\DashArrowLine(230,50)(250,50){1}
  \Photon(250,50)(250,80){-2}{4} 
\DashArrowLine(250,50)(310,50){1}
\DashArrowLine(310,50)(330,50){1}
\GlueArc(280,50)(30,180,360){3}{10.5} 
\Vertex(250,50){1}\Text(255,55)[lb]{$e_r t_r$}
\Vertex(310,50){1}\Text(310,55)[b]{$t_r$}
\Text(280,-5)[]{(b)} 

\DashArrowLine(360,50)(380,50){1}
\DashArrowLine(380,50)(440,50){1}
  \Photon(440,50)(440,80){2}{4} 
\DashArrowLine(440,50)(460,50){1}
\GlueArc(410,50)(30,180,360){3}{10.5} 
\Vertex(380,50){1}\Text(380,55)[b]{$t_r$}
\Vertex(440,50){1}\Text(435,55)[rb]{$e_r t_r$}
\Text(410,-5)[]{(c)}  
\end{picture} 
\end{center}
\begin{center}
\begin{picture}(500,100)(100,-15)
\DashArrowLine(100,50)(130,50){1}
  \Photon(130,50)(130,80){2}{4}
\DashArrowLine(130,50)(160,50){1}
\GlueArc(130,30)(18,90,450){3}{15}
\Vertex(130,50){1}\Text(135,55)[lb]{$e_r t_r^2$}
\Text(130,-5)[]{(d)} 

\DashArrowLine(190,20)(210,20){1}
\DashArrowLine(210,20)(270,20){1}
\DashArrowLine(270,20)(290,20){1}
\GlueArc(240,20)(30,0,90){3}{5} 
\Vertex(270,20){1}\Text(270,15)[t]{$t_r$}
  \Photon(240,50)(240,80){2}{4}
  \Vertex(240,50){1}\Text(245,55)[lb]{$t_r$}
\PhotonArc(240,20)(30,90,180){-2}{6}\Text(210,15)[t]{$e_r$}
\Text(240,-5)[]{(e)} 

\DashArrowLine(320,20)(340,20){1}
\DashArrowLine(340,20)(400,20){1}
\DashArrowLine(400,20)(420,20){1}
\PhotonArc(370,20)(30,0,90){2}{6}\Text(400,15)[t]{$e_r$}
  \Photon(370,50)(370,80){-2}{4}
  \Vertex(370,50){1}\Text(375,55)[lb]{$t_r$}
\GlueArc(370,20)(30,90,180){3}{5} 
\Vertex(340,20){1}\Text(340,15)[t]{$t_r$}
\Text(370,-5)[]{(f)}  

\DashArrowLine(450,20)(480,20){1}
\DashArrowLine(480,20)(510,20){1}
\GlueArc(480,40)(18,-90,90){3}{7.5}
    \Photon(480,58)(480,88){2}{4}
    \Vertex(480,58){1}\Text(485,63)[lb]{$t_r$}
\PhotonArc(480,40)(18,90,270){-2}{8}
\Vertex(480,20){1}\Text(481,15)[t]{$e_r t_r$}
\Text(480,-5)[]{(g)} 
\end{picture} \\ 
Fig. 6. $O(t_r^2)$ corrections to $Z_1$.
\end{center} 

The Ward-Takahashi identity is described by an equation in which 
the QED vertex function, $\Lambda_{\mu}$, is equivalent to 
a derivative of the fermion self-energy, $\Sigma$, with respect to 
the external fermion momentum. 
Now, the vertex of type $h^n\!-\!{\bar \psi}\!-\!\psi$ 
contains derivatives of fermion fields, so that we must take care 
when determining the momentum dependence at the vertex. 
Let us first consider diagram (a) in Fig. 5. 
The fermion self-energy function from this diagram in the momentum space 
has, apart from an overall coefficient, the form  
\bb
    \Sigma^{(a)} =  \gm_{\mu}\fr{1}{k\!\!\!/+p\!\!\!/}\gm_{\lam} 
                 \fr{1}{k^4}(I_H)_{\mu\nu,\lam\s}
                 (k+2p)_{\nu}(k+2p)_{\s},  
\ee
where $p$ is the external fermion momentum and $k$ is the 
internal traceless-mode momentum. The momentum dependence, $k+2p$, 
comes from the vertex of type $h\!-\!{\bar \psi}\!-\!\psi$. 
Thus, the derivative of $\Sigma$ with respect to $p$ acts not only on 
the fermion propagator but also on the momentum factor at the 
vertices, and hence  
\bba
    \fr{\pd}{\pd p_{\rho}} \Sigma^{(a)}  
    &=& -\gm_{\mu}\fr{1}{k\!\!\!/+p\!\!\!/}\gm_{\rho} 
        \fr{1}{k\!\!\!/+p\!\!\!/}\gm_{\lam}
              \fr{1}{k^4}(I_H)_{\mu\nu,\lam\s}
                 (k+2p)_{\nu}(k+2p)_{\s}
             \nonumber  \\ 
    &&  + \gm_{\mu}\fr{1}{k\!\!\!/+p\!\!\!/}\gm_{\lam} 
            \fr{1}{k^4}(I_H)_{\mu\nu,\lam\s}
             \bigl\{  2\dl_{\nu\rho} (k+2p)_{\s} 
                     +2\dl_{\s\rho} (k+2p)_{\nu} \bigr\}.
\eea
The first term on the r.h.s. of this relation corresponds to the vertex 
function $\Lambda_{\rho}^{(a)}$ from diagram (a) in Fig. 6 
when evaluated at zero momentum for the external photon.  
The second is the sum of $\Lambda_{\rho}^{(b)}$ and $\Lambda_{\rho}^{(c)}$
from diagrams (b) and (c) in Fig. 6, respectively. Thus, the sum of the  
vertex functions from (a) to (c) in Fig. 6 is equivalent to 
the derivative of $\Sigma^{(a)}$ with respect to $p$. 
Hence, we can show that $Z_2$ from diagram (a) in Fig. 5 
is equivalent to the sum of $Z_1$ from diagrams (a) to (c) in Fig. 6.  

We can easily show that $Z_2$ from diagram (b) in Fig. 5 is 
equivalent to $Z_1$ from diagram (d) in Fig. 6. 
In general, the Ward-Takahashi identity holds  for 
all QED vertex functions obtained by attaching an external photon 
in all possible ways to a fermion line carrying the external charge 
flow, as in diagrams (a) to (d) in Fig. 6.
This is because the vertex ${\bar \psi}\gm_{\mu} \psi A_{\nu} (h^n)_{\mu\nu}$  
is given through the minimal 
substitution $p_{\nu} \arr p_{\nu} -e A_{\nu}$ from the vertex 
${\bar \psi}\gm_{\mu}p_{\nu} \psi (h^n)_{\mu\nu}$, apart from 
vertices of the type $h^n\!-\!{\bar \psi}\!-\!\psi$ 
with a derivative of the traceless mode, which is not affected by the   
differentiation with respect to the external fermion momentum.   

For the diagrams (e) to (g) in Fig. 6, there is no associated 
fermion self-energy diagram. 
However, such diagrams do not have divergences, 
because a vertex of the type $h\!-\!A\!-\!A$ contains  
derivatives of the photon fields such that these diagrams vanish when 
evaluated at zero momentum for the external photon. 
Thus, we can show $Z_1=Z_2$ at $O(t_r^2)$. 
In general, QED vertex corrections with an external photon that stems from 
the gravitational vertices of types $h^n \!-\!A\!-\!A$ and 
$\phi^n \!-\!A\!-\!A$ vanish when evaluated at zero momentum.  

To see that $Z_1=Z_2$ holds at higher orders, 
we need to show that all QED vertex corrections obtained by 
attaching an external photon to internal fermion loops 
vanish when evaluated at zero momentum for the external photon. 
To see this, we need to show that Furry's theorem holds in the case 
including gravitational fields, that is, that a fermion loop with an odd number 
of QED vertices vanishes even though arbitrary number of the traceless mode 
is attached on the fermion loop. This can be easily shown by noting that 
the photon field changes sign under charge conjugation, 
but the gravitational field does not change sign.  
Thus, QED vertex corrections including a fermion loop with 
an odd number of QED vertices, such as diagram (a) in Fig. 7, vanish.   

There are diagrams that are not forbidden by the generalized 
form of Furry's theorem, such as diagram (b) in Fig. 7. 
However, such diagrams vanish due to a gauge invariance 
when evaluated at zero momentum for the external 
photon.     
Thus, the identity $Z_1=Z_2$ holds also at higher orders.  
\begin{center}
\begin{picture}(400,100)(50,-15)
\DashArrowLine(100,20)(120,20){1}
\DashArrowLine(120,20)(180,20){1}
\DashArrowLine(180,20)(200,20){1}
\GlueArc(150,20)(30,0,71){3}{4} 
  \Vertex(180,20){1} 
  \Vertex(160,49){1} 
  \DashArrowArc(150,50)(10,-135,225){1}
    \Photon(150,60)(150,80){2}{3}
      \Gluon(150,40)(140,20){-3}{3}
      \Photon(150,40)(160,20){2}{3.5}
         \Vertex(150,40){1}\Vertex(140,20){1}
\PhotonArc(150,20)(30,109,180){-2}{5} 
\Text(150,-5)[]{(a)}   

\DashArrowLine(300,20)(320,20){1}
\DashArrowLine(320,20)(380,20){1}
\DashArrowLine(380,20)(400,20){1}
\GlueArc(350,20)(30,0,71){3}{4} 
  \Vertex(380,20){1} 
  \Vertex(360,49){1} 
  \DashArrowArc(350,50)(10,-135,225){1}
  \Photon(350,60)(350,80){2}{3}
     \Gluon(350,40)(340,20){-3}{3}
     \Gluon(350,40)(360,20){3}{3} 
       \Vertex(350,40){1}\Vertex(340,20){1}\Vertex(360,20){1}
\PhotonArc(350,20)(30,109,180){-2}{5} 
\Text(350,-5)[]{(b)} 
\end{picture} \\ 
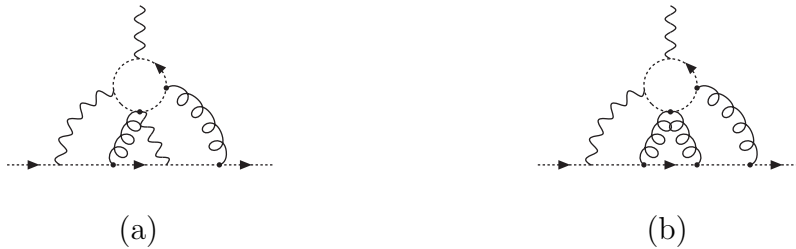
Fig. 7. Examples of QED vertex corrections obtained by 
attaching an external photon to an internal fermion loop. 
\end{center} 

\subsection{Gravitational corrections to $Z_3$}
\noindent

The gravitational corrections to $Z_3$ at $O(t_r^2)$ are given by 
two diagrams, (a) and (b) in Fig. 8. 
The UV divergences of these two diagrams exactly cancel, so that there 
is no contribution to $Z_3$ at $O(t_r^2)$.   
\begin{center}
\begin{picture}(400,100)(0,0)
\Photon(100,50)(120,50){2}{3}
  \Vertex(120,50){1}\Text(115,45)[rt]{$t_r$}
  \GlueArc(140,50)(20,0,180){3}{6}
  \PhotonArc(140,50)(20,180,360){-2}{9.5}
\Photon(160,50)(180,50){2}{3}
  \Vertex(160,50){1}\Text(165,45)[lt]{$t_r$}
\Text(140,10)[]{(a)}

\Photon(230,40)(290,40){2}{7}
\GlueArc(260,60)(20,-90,270){3}{15}
  \Vertex(260,40){1}\Text(260,35)[t]{$t_r^2$}
\Text(260,10)[]{(b)}
\end{picture} \\
Fig. 8. $O(t_r^2)$ corrections to $Z_3$. 
\end{center}

At higher order, there is a contribution to $Z_3$ from gravity. 
Here, we calculate diagrams including an internal conformal mode. 
Such contributions appear at $O(e_r^4)$. The diagrams are depicted in 
Fig. 9. They all yield simple poles.  
Here, diagrams including the ordinary two-loop 
photon self-energy and the four-photon interaction  
as a subdiagram are not depicted in Fig. 9, because each sum of 
such diagrams is finite, as discussed in Subsection 6.2. 
{}From a simple counting of the number of loops and $\eps$'s at the 
vertices, diagrams including $\phi^2 \!-\! A \!-\! A$ vertices  
and diagrams including two internal conformal-mode lines are  
trivially finite at $O(e_r^4)$, and therefore they are also omitted.  
\begin{center}
\begin{picture}(500,100)(70,0)
\Photon(100,50)(120,50){2}{3}
  \Vertex(120,50){1}\Text(115,45)[rt]{$n_F e_r^2$}
\CArc(140,36)(25,33,147)
  \PhotonArc(140,50)(20,180,360){-2}{9.5}
\Photon(160,50)(180,50){2}{3}
  \Vertex(160,50){1}\Text(165,45)[lt]{$n_F e_r^2$}
\Text(140,10)[]{(a)}

\Photon(210,50)(230,50){2}{3}
  \Vertex(230,50){1}\Text(225,45)[rt]{$\eps$}
\CArc(250,36)(25,33,147)
  \PhotonArc(250,50)(20,180,247){-2}{4}
     \DashArrowArc(250,30)(8,-90,270){1}
  \PhotonArc(250,50)(20,293,360){2}{4}
\Photon(270,50)(290,50){2}{3}
  \Vertex(270,50){1}\Text(275,45)[lt]{$n_F e_r^2$}
\Text(250,10)[]{(b)}

\Photon(320,50)(340,50){2}{3}
  \Vertex(340,50){1}\Text(335,45)[rt]{$\eps$}
\CArc(365,34)(30,31,149)
  \PhotonArc(365,50)(25,180,212){-2}{2}
   \DashArrowArc(350,33)(7,-120,240){1}
  \PhotonArc(365,50)(25,246,294){2}{3} 
   \DashArrowArc(380,33)(7,-60,300){1}
  \PhotonArc(365,50)(25,328,360){2}{2}
\Photon(390,50)(410,50){2}{3}
  \Vertex(390,50){1}\Text(395,45)[lt]{$\eps$}
\Text(365,10)[]{(c)}
\end{picture} \\
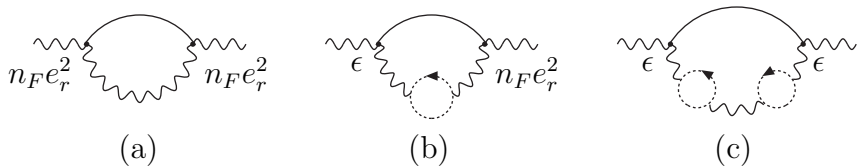
Fig. 9. $O(e_r^4)$ diagrams yielding simple poles contributing to $Z_3$. 
\end{center}
 
We also calculate the residue of the double-pole term of $Z_3$ 
at $O(e_r^6)$. It produces a simple pole counterterm of the 
type $\phi \!-\!A\!-\!A$, as discussed in Section 4, 
which is used to show the finiteness of the vertex corrections below. 
There are only three diagrams contributing to double poles, 
which are depicted in Fig. 10.  
\begin{center}
\begin{picture}(500,100)(70,-10)
\Photon(100,50)(120,50){2}{3}
  \Vertex(120,50){1}\Text(115,45)[rt]{$n_F e_r^2$}
\CArc(140,36)(25,33,147)
  \PhotonArc(140,50)(20,180,247){-2}{4}
     \DashArrowArc(140,30)(8,-90,270){1}
  \PhotonArc(140,50)(20,293,360){2}{4}
\Photon(160,50)(180,50){2}{3}
  \Vertex(160,50){1}\Text(165,45)[lt]{$n_F e_r^2$}
\Text(140,0)[]{(a)}

\Photon(210,50)(230,50){2}{3}
  \Vertex(230,50){1}\Text(225,45)[rt]{$\eps$}
\CArc(255,34)(30,31,149)
  \PhotonArc(255,50)(25,180,212){-2}{2}
   \DashArrowArc(240,33)(7,-120,240){1}
  \PhotonArc(255,50)(25,246,294){2}{3} 
   \DashArrowArc(270,33)(7,-60,300){1}
  \PhotonArc(255,50)(25,328,360){2}{2}
\Photon(280,50)(300,50){2}{3}
  \Vertex(280,50){1}\Text(285,45)[lt]{$n_F e_r^2$}
\Text(255,0)[]{(b)}

\Photon(330,50)(350,50){2}{3}
  \Vertex(350,50){1}\Text(345,45)[rt]{$\eps$}
\CArc(380,33)(35,30,150)
  \PhotonArc(380,50)(30,180,209){-2}{2.5}
     \DashArrowArc(357,30)(6,-135,255){1}
  \PhotonArc(380,50)(30,231,259){2}{2.5} 
     \DashArrowArc(380,20)(6,-90,270){1}
  \PhotonArc(380,50)(30,281,309){2}{2.5}
     \DashArrowArc(403,30)(6,-45,315){1}
  \PhotonArc(380,50)(30,331,360){-2}{2.5}
\Photon(410,50)(430,50){2}{3}
  \Vertex(410,50){1}\Text(415,45)[lt]{$\eps$}
\Text(380,0)[]{(c)}
\end{picture} \\
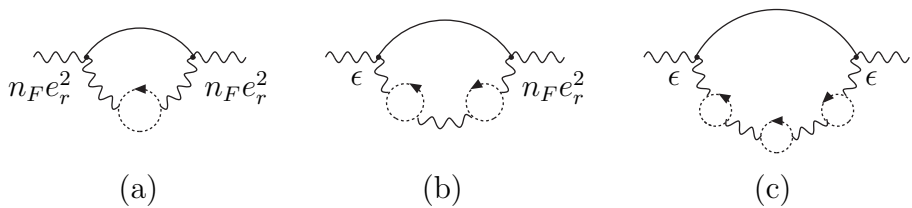
Fig. 10. $O(e_r^6)$ diagrams yielding double poles contributing to $Z_3$. 
\end{center}

In summary, we obtain 
\bba
   && Z_3 = 1 -\fr{4n_F}{3}\fr{\te_r^2}{(4\pi)^2}\fr{1}{\eps} 
          + \biggl( -2n_F +\fr{8}{27}\fr{n_F^2}{\tb_c} \biggr) 
               \fr{\te_r^4}{(4\pi)^4}\fr{1}{\eps} 
                 \nonumber \\ 
   && \qquad\qquad
          +\biggl( -\fr{8n_F^2}{9} + \fr{8}{81}\fr{n_F^3}{\tb_c} \biggr) 
               \fr{\te_r^6}{(4\pi)^6}\fr{1}{\eps^2} 
          + O(\te_r^2 \tt_r^2, ~\tt_r^4). 
                \label{Z3}
\eea
Here, the simple-pole and the double-pole terms proportional 
to $1/\tb_c$ are the contributions from diagrams in Figs. 9 and 10, 
respectively.  The other terms are well-known QED corrections~\cite{rr}. 

We now calculate the beta function for the QED coupling constant, 
\bb
    \b_e = \mu \fr{d}{d \mu} \te_r. 
          \label{beta-e1}
\ee 
The renormalization group equation (RGE) is defined as an equation 
stipulating that the bare quantities are independent of the arbitrary mass scale,  
$\mu \fr{d}{d \mu} ( \hbox{bare actions} )=0$. 
Thus, we obtain the equation $\mu \fr{d}{d \mu} e=0$, so that 
the beta function can be expressed as 
\bb
    \b_e=-\eps \te_r +\fr{\te_r}{2}\fr{\mu}{Z_3}\fr{dZ_3}{d\mu}. 
            \label{beta-e2}
\ee
The bare constant $b$ is treated in a slightly 
different way. Since $\mu \fr{d}{d \mu} b=0$, 
the dimensionless quantity $\tb$ satisfies $\mu \fr{d}{d \mu} \tb= 2\eps \tb$. 
We can re-expand $\tb$     
using the dimensionless bare coupling constants $\te$ and $\tt$ as 
$\tb=\sum_n \fr{\tb^{\pp}_n}{(D-4)^n}$, where
\bb
    \tb^{\pp}_n = \sum_{r,s \geq 0} \tb_{c,n}^{\pp (r,s)}\te^{2r}\tt^{2s}, 
\ee
and hence $\tb_{c,1}^{\pp (0,0)}=\fr{\tb_c}{(4\pi)^2}$. 
The RGE implies that the coefficients $\tb_{c,n}^{\pp (r,s)}$ are 
not running coupling constants and that they satisfy the linear equation   
\bb
    \mu \fr{d}{d \mu} \tb_{c,n}^{\pp (r,s)} 
    = 2(1+r+s)\eps \tb_{c,n}^{\pp (r,s)}. 
            \label{bcn}
\ee 
Thus, if we wish to derive the beta function from information 
concerning the poles of the renormalization factor, we must consider 
$\tb_{c,n}^{\pp (r,s)}$ as unknown constants and use 
equation (\ref{bcn}). 
After deriving finite expressions, we must substitute the values 
of $\tb_{c,n}^{\pp (r,s)}$ into them. 
 
We can now calculate the beta function for the QED coupling constant. 
Here, the related coefficient is only $\tb_c$, which satisfies 
$\mu \fr{d}{d \mu} \tb_c= 2\eps \tb_c$. 
For the moment, consider the general expression of $Z_3$ 
expanded in $\te_r$ and the inverse of $\tb_c$ as   
\bb
   Z_3=1 + \fr{A_1}{\eps} +\fr{A_2}{\eps^2}  +\cdots
         + \fr{1}{\tb_c} \biggl( \fr{B_1}{\eps} 
               +\fr{B_2}{\eps^2} +\cdots \biggr) 
         + \fr{1}{\tb_c^2} \biggl( \fr{C_1}{\eps} 
               +\fr{C_2}{\eps^2} +\cdots \biggr) 
         +\cdots,
\ee  
where
\bba
   &&  A_1=\sum_{n \geq 1}A_{1,n}\te_r^{2n}, \quad 
       A_2=\sum_{n \geq 3}A_{2,n}\te_r^{2n},
             \nonumber     \\
   &&  B_1=\sum_{n \geq 2}B_{1,n}\te_r^{2n}, \quad 
       B_2=\sum_{n \geq 3}B_{2,n}\te_r^{2n},
                  \\ 
   &&  C_1=\sum_{n \geq 3}C_{1,n}\te_r^{2n}, \quad 
       C_2=\sum_{n \geq 4}C_{2,n}\te_r^{2n}.
            \nonumber 
\eea
Solving the simultaneous equations (\ref{beta-e1}) and 
(\ref{beta-e2}) order by order in $\eps$, 
we obtain the relations between the residues of 
double poles and simple poles  
\bb
      A_{2,3}=-\fr{1}{3}A_{1,1}A_{1,2}, \quad 
      B_{2,3}=-\fr{1}{4}A_{1,1}B_{1,2},  
            \label{pole-relation}
\ee
and a finite expression for the beta function in the $\eps \arr 0$ limit,
\bb
  \b_e = -A_{1,1}e_r^3 -2A_{1,2}e_r^5 -3A_{1,3}e_r^7 
         -3 B_{1,2}\fr{e_r^5}{b_c} -4 B_{1,3}\fr{e_r^7}{b_c}  
         -5 C_{1,3}\fr{e_r^7}{b_c^2} +\cdots. 
\ee
The residues summarized by (\ref{Z3}) do satisfy relation 
(\ref{pole-relation}). 
Note that the coefficient, $3$, in front of the 
$B_{1,2}$ term implies that the diagrams in Fig. 9 are 
essentially of the three-loop kind. 
This is reasonable, because the vertex of 
type $\phi \!-\!A\!-\!A$ with $\e_r^2$  
essentially corresponds to a one-loop diagram. 
Thus, the treatment for $\tb$ mentioned above may be considered 
a trick to count the number of loops correctly.   

Thus, we obtain 
\bb
    \b_e = \fr{4n_F}{3}\fr{e_r^3}{(4\pi)^2} 
           + 4 n_F \fr{e_r^5}{(4\pi)^4} 
           - \fr{8}{9}\fr{n_F^2}{b_c}\fr{e_r^5}{(4\pi)^4} 
           + O(e_r^3 t_r^2, e_r t_r^4). 
               \label{beta-e}
\ee 
Substituting the value of $b_c$ into (\ref{bcv}), we can see that the 
sign of the $O(e_r^5)$ term becomes negative for $n_F \geq 24$. 
This result, as well as the result for $\b_t$, (\ref{beta-t}), suggests 
that all gravitational corrections give negative contributions 
to the beta functions. Hence, provided that the residues of the simple poles 
of $Z_3$ at $O(e_r^2 t_r^2)$ and $O(t_r^4)$ are positive, 
which give negative contributions to $\b_e$, 
we can consider the scenario in which the UV fixed point of the QED coupling 
constant becomes zero, or close to zero, asymptotically beyond the Planck 
mass scale.

\subsection{$O(e_r^6)$ renormalization of $\phi \!-\!A\!-\!A$ vertex I}
\noindent

The renormalization factor $Z_3$ has double poles at $O(e_r^6)$.  
This means that a new counterterm for 
the vertex of type $\phi \!-\!A\!-\!A$ appears 
at $O(e_r^6)$. In the following two subsections, we directly check 
whether this vertex is renormalizable using only the data 
for $Z_3$ and $Z_{\phi}=1$. 

In this subsection, we first consider the case in which there are no 
internal gravitational fields. The diagrams that potentially have  
UV divergences at $O(e_r^6)$ are divided into four groups, as depicted  
in Fig. 11. These are described in the following.    
\begin{center}
\begin{picture}(570,100)(70,0)
\Photon(100,50)(170,50){2}{10}
\Line(135,50)(135,80)
\GCirc(135,50){15}{1}
\Text(135,50)[]{$2$}
\Text(135,10)[]{(a)}

\Photon(200,50)(270,50){2}{10}
\Line(235,50)(235,85)
\GCirc(235,50){18}{1}
\Text(235,55)[]{$3$}\Text(235,43)[]{$(S)$}
\Text(235,10)[]{(b)}

\Photon(300,50)(370,50){2}{10}
\Line(335,50)(335,85)
\GCirc(335,50){18}{1}
\Text(335,55)[]{$3$}\Text(335,43)[]{$(D)$}
\Text(335,10)[]{(c)}

\Photon(400,50)(440,50){2}{6}
\Line(420,50)(420,70)
\GCirc(420,50){4}{1}
  \Line(417,53)(423,47)
  \Line(417,47)(423,53)
\Text(420,10)[]{(d)}
\end{picture} \\
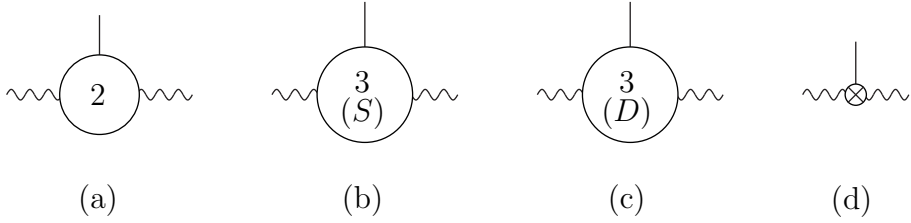
Fig. 11. $O(e_r^6)$ $\phi \!-\!A\!-\!A$ vertex diagrams  
without internal gravitational fields. 
\end{center}

Diagram (a) in Fig. 11 represents the sum of diagrams 
given by attaching a line with $e_r^2$ 
at the end to internal photons in the ordinary two-loop photon 
self-energy (2LPSE) diagrams [Fig. 11-(a)].
Since the sum of the diagrams for 2LPSE gives only a simple pole, 
diagram Fig.11-(a) also gives a simple pole.  
\begin{center}
\begin{picture}(500,100)(70,0)
\Photon(100,50)(170,50){2}{10}
\Line(135,50)(135,80)
\GCirc(135,50){15}{1}
\Text(135,50)[]{$2$}

\Text(185,50)[]{$=$}

\Photon(200,50)(220,50){2}{3}
\DashArrowArc(250,50)(30,0,180){1}
     \DashArrowArc(250,50)(30,180,360){1}
  \PhotonArc(250,80)(30,210,330){2}{8}     
    \Line(250,50)(250,65) 
    \Vertex(250,50){1}\Text(250,40)[]{$n_F e_r^2$}
\Photon(280,50)(300,50){2}{3} 

\Text(315,50)[]{$+$}  

\Photon(330,50)(350,50){2}{3}
\DashArrowArc(380,50)(30,30,210){1}
     \DashArrowArc(380,50)(30,210,390){1}
  \Photon(380,80)(380,20){2}{7}     
    \Line(380,50)(395,50) 
    \Vertex(380,50){1}\Text(378,50)[r]{$n_F e_r^2$}
\Photon(410,50)(430,50){2}{3} 
\end{picture} \\
Fig. 11-(a).  $\phi \!-\!A\!-\!A$ vertex diagrams $(a)$. 
\end{center}

Diagram (b) in Fig. 11 represents the sum of diagrams 
given by attaching a line with $\eps$ 
at the end to internal photons in the ordinary three-loop photon 
self-energy (3LPSE) with a single fermion loop [Fig. 11-(b)]. 
It is known that the sum of the diagrams for 3LPSE with a single 
fermion loop yields only a simple pole, and thus  
diagram Fig. 11-(b) becomes finite, due to the $\eps$ at the vertex. 
\begin{center}
\begin{picture}(500,100)(70,0)
\Photon(100,50)(170,50){2}{10}
\Line(135,50)(135,85)
\GCirc(135,50){18}{1}
\Text(135,55)[]{$3$}\Text(135,43)[]{$(S)$}

\Text(185,50)[]{$=$}

\Photon(200,50)(220,50){2}{3}
\DashArrowArc(250,50)(30,0,180){1}
     \DashArrowArc(250,50)(30,180,360){1}
  \PhotonArc(250,90)(30,222,318){2}{5.5}     
    \Line(250,62)(250,73) 
    \Vertex(250,62){1}\Text(250,56)[]{$\eps$}
  \PhotonArc(250,10)(30,42,138){2}{5.5}
\Photon(280,50)(300,50){2}{3} 

\Text(315,50)[]{$+$}  

\Photon(330,50)(350,50){2}{3}
\DashArrowArc(380,50)(30,0,180){1}
     \DashArrowArc(380,50)(30,180,360){1}
  \Photon(367,77)(367,23){2}{6.5}     
    \Line(369,50)(382,50) 
    \Vertex(369,50){1}\Text(362,50)[]{$\eps$}
  \Photon(393,77)(393,23){2}{6.5}
\Photon(410,50)(430,50){2}{3} 

\Text(450,50)[]{$+$}\Text(470,50)[]{$\cdots$}
\end{picture} \\
Fig. 11-(b).  $\phi \!-\!A\!-\!A$ vertex diagrams $(b)$. 
\end{center}

Diagram (c) in Fig. 11 represents the sum of diagrams 
given by attaching a line with $\eps$ 
at the end to internal photons in 3LPSE 
with two fermion loops [Fig. 11-(c)]. 
Since the sum of the diagrams for 3LPSE with two 
fermion loops gives a double pole, but not a triple pole,  
diagram Fig. 11-(c) yields a simple pole. 
\begin{center}
\begin{picture}(500,100)(70,0)
\Photon(100,50)(170,50){2}{10}
\Line(135,50)(135,85)
\GCirc(135,50){18}{1}
\Text(135,55)[]{$3$}\Text(135,43)[]{$(D)$}

\Text(185,50)[]{$=$}

\Photon(200,50)(220,50){2}{3}
\DashArrowArc(250,50)(30,0,180){1}
     \DashArrowArc(250,50)(30,180,360){1}
  \PhotonArc(250,80)(30,210,255){-2}{3}     
     \DashArrowArc(250,50)(8,-90,270){1}
  \PhotonArc(250,80)(30,285,330){2}{3}
    \Line(235,54)(235,70) 
    \Vertex(235,54){1}\Text(234,52)[rt]{$\eps$}
\Photon(280,50)(300,50){2}{3} 

\Text(315,50)[]{$+$}  

\Photon(330,50)(350,50){2}{3}
\DashArrowArc(380,50)(30,30,210){1}
     \DashArrowArc(380,50)(30,210,390){1}
  \Photon(380,80)(380,58){2}{3}     
  \DashArrowArc(380,50)(8,-180,180){1}
  \Photon(380,42)(380,20){2}{3}
    \Line(378,67)(395,67) 
    \Vertex(378,67){1}\Text(375,67)[]{$\eps$}
\Photon(410,50)(430,50){2}{3} 
\end{picture} \\
Fig. 11-(c).  $\phi \!-\!A\!-\!A$ vertex diagrams $(c)$. 
\end{center}

Diagram (d) in Fig. 11 is the $O(e_r^6)$ direct counterterm 
mentioned above. 
The sum of all UV divergences  evaluated 
at zero momentum for the external conformal mode 
vanishes exactly:
\bb
    \Gamma^{\phi AA}_{\mu\nu}(0;k,-k) \vert_I
    = \biggl\{ -\fr{8}{3}+\fr{16}{9}+\fr{8}{9} \biggr\} n_F^2
       \fr{e_r^6}{(4\pi)^6}\fr{1}{\eps} 
       \bigl( \dl_{\mu\nu} k^2 -k_{\mu}k_{\nu} \bigr) =0. 
              \label{GmI}
\ee
We here normalize the effective action in the momentum representation 
as $\Gamma = \int  [dp][dk][dl](2\pi)^D \dl^D (p+k+l)
\phi(p)A^r_{\mu}(k)A^r_{\nu}(l) \Gamma^{\phi AA}_{\mu\nu}(p;k,l)$, 
where $[dp]=\fr{d^D p}{(2\pi)^D}$. 
The first and second terms in the braces result from the diagrams  
in Figs. 11-(a) and 11-(c), respectively.   
The third term in (\ref{GmI}) comes from the direct counterterm 
diagram (d) in Fig. 11 related to the double pole term in $Z_3$. 

\subsection{$O(e_r^6)$ renormalization of $\phi \!-\!A\!-\!A$ vertex II}
\noindent

Next, we consider the vertex diagrams of  
type $\phi \!-\! A \!-\! A$, including internal conformal-mode lines
that yield divergences proportional to $e_r^6/b_c$. 
The diagrams are depicted in Fig. 12, where,   
for diagrams (e) to (g), the other assignments of $\eps$ and $\e_r^2$ 
on the vertices should be taken into account. 
As mentioned above, the sum of diagrams at $O(e_r^6)$ 
including 2LPSE as a subdiagram is finite. 
The same conclusion holds for diagrams including 
3LPSE, four- and six-photon interactions as a subdiagram. 
Therefore, none of them are depicted in Fig. 12.  
The new features that do not appear in the previous cases are that there are 
contributions from vertices of types $\phi^3$ and 
$\phi^2 \!-\! A\!-\!A$. 

\begin{center}
\begin{picture}(500,100)(70,0)
\Photon(100,50)(120,50){2}{3}
  \Vertex(120,50){1}\Text(115,45)[rt]{$n_F e_r^2$}
\CArc(140,36)(25,33,147)
    \Line(140,61)(140,78)
      \Vertex(140,61){1}\Text(140,57)[t]{$\eps b_1$}
  \PhotonArc(140,50)(20,180,247){-2}{4}
     \DashArrowArc(140,30)(8,-90,270){1}
  \PhotonArc(140,50)(20,293,360){2}{4}
\Photon(160,50)(180,50){2}{3}
  \Vertex(160,50){1}\Text(165,45)[lt]{$n_F e_r^2$}
\Text(140,0)[]{(a)}

\Photon(210,50)(230,50){2}{3}
  \Vertex(230,50){1}\Text(225,45)[rt]{$\eps$}
\CArc(255,34)(30,31,149)
  \Line(255,64)(255,81)
    \Vertex(255,64){1}\Text(255,62)[t]{$\eps b_1$}
  \PhotonArc(255,50)(25,180,212){-2}{2}
   \DashArrowArc(240,33)(7,-120,240){1}
  \PhotonArc(255,50)(25,246,294){2}{3} 
   \DashArrowArc(270,33)(7,-60,300){1}
  \PhotonArc(255,50)(25,328,360){2}{2}
\Photon(280,50)(300,50){2}{3}
  \Vertex(280,50){1}\Text(285,45)[lt]{$n_F e_r^2$}
\Text(255,0)[]{(b)}

\Photon(330,50)(350,50){2}{3}
  \Vertex(350,50){1}\Text(345,45)[rt]{$\eps$}
\CArc(380,33)(35,30,150)
  \Line(380,68)(380,85)
    \Vertex(380,68){1}\Text(380,65)[t]{$\eps b_1$}
  \PhotonArc(380,50)(30,180,209){-2}{2.5}
     \DashArrowArc(357,30)(6,-135,255){1}
  \PhotonArc(380,50)(30,231,259){2}{2.5} 
     \DashArrowArc(380,20)(6,-90,270){1}
  \PhotonArc(380,50)(30,281,309){2}{2.5}
     \DashArrowArc(403,30)(6,-45,315){1}
  \PhotonArc(380,50)(30,331,360){-2}{2.5}
\Photon(410,50)(430,50){2}{3}
  \Vertex(410,50){1}\Text(415,45)[lt]{$\eps$}
\Text(380,0)[]{(c)}
\end{picture} 
\end{center}
\begin{center}
\begin{picture}(500,100)(70,20)
\Photon(100,50)(120,50){2}{3}
  \Vertex(120,50){1}\Text(120,45)[t]{$n_F e_r^2$}
\PhotonArc(140,50)(20,0,180){-2}{8.5}
    \Line(140,68)(140,85)
      \Vertex(140,68){1}\Text(144,75)[lb]{$n_F e_r^2$}
\Line(120,50)(160,50)
\Photon(160,50)(180,50){2}{3}
  \Vertex(160,50){1}\Text(160,45)[t]{$n_F e_r^2$}
\Text(140,20)[]{(d)}

\Photon(210,50)(230,50){2}{3}
  \Vertex(230,50){1}\Text(230,45)[t]{$n_F e_r^2$}
\PhotonArc(255,50)(25,0,25){-2}{1.5}
   \DashArrowArc(273,66)(7,-135,225){1}
\PhotonArc(255,50)(25,58,180){2}{7}
    \Line(255,73)(255,90)
      \Vertex(255,73){1}\Text(255,70)[t]{$\eps$}
\Line(230,50)(280,50)
\Photon(280,50)(300,50){2}{3}
  \Vertex(280,50){1}\Text(280,45)[t]{$n_F e_r^2$}
\Text(255,20)[]{(e)}

\Photon(330,50)(350,50){2}{3}
  \Vertex(350,50){1}\Text(350,45)[t]{$\eps$}
\PhotonArc(380,50)(30,0,14){-2}{1}
   \DashArrowArc(407,63)(6,-135,225){1}
\PhotonArc(380,50)(30,38,55){2}{1.5}
   \DashArrowArc(392,77)(6,-120,240){1}
\PhotonArc(380,50)(30,78,180){2}{7}
    \Line(380,78)(380,95)
      \Vertex(380,78){1}\Text(380,75)[t]{$\eps$}
\Line(350,50)(410,50)
\Photon(410,50)(430,50){2}{3}
  \Vertex(410,50){1}\Text(410,45)[t]{$n_F e_r^2$}
\Text(380,20)[]{(f)}
\end{picture} 
\end{center}
\begin{center}
\begin{picture}(500,100)(70,20)
\Photon(100,50)(120,50){2}{3}
  \Vertex(120,50){1}\Text(120,45)[t]{$\eps$}
\PhotonArc(145,50)(25,0,25){-2}{1.5}
   \DashArrowArc(163,66)(7,-135,225){1}
\PhotonArc(145,50)(25,59,121){2}{4}
   \DashArrowArc(127,66)(7,-45,315){1}
\PhotonArc(145,50)(25,155,180){-2}{1.5}
    \Line(145,75)(145,90)
      \Vertex(145,75){1}\Text(145,70)[t]{$\eps$}
\Line(120,50)(170,50)
\Photon(170,50)(190,50){2}{3}
  \Vertex(170,50){1}\Text(170,45)[t]{$n_F e_r^2$}
\Text(145,20)[]{(g)}

\Photon(220,50)(240,50){2}{3}
  \Vertex(240,50){1}\Text(240,45)[t]{$\eps$}
\PhotonArc(275,50)(35,0,12){-2}{1}
   \DashArrowArc(308,62)(5,-135,225){1}
\PhotonArc(275,50)(35,28,39){2}{1}
   \DashArrowArc(299,76)(5,-120,240){1}
\PhotonArc(275,50)(35,56,65){2}{1}
   \DashArrowArc(285,84)(5,-120,240){1}
\PhotonArc(275,50)(35,82,180){2}{8}
    \Line(275,83)(275,100)
      \Vertex(275,83){1}\Text(275,80)[t]{$\eps$}
\Line(240,50)(310,50)
\Photon(310,50)(330,50){2}{3}
  \Vertex(310,50){1}\Text(310,45)[t]{$\eps$}
\Text(275,20)[]{(h)}

\Photon(360,50)(380,50){2}{3}
  \Vertex(380,50){1}\Text(380,45)[t]{$\eps$}
\PhotonArc(410,50)(30,0,14){-2}{1}
   \DashArrowArc(437,63)(6,-135,225){1}
\PhotonArc(410,50)(30,38,55){2}{1.5}
   \DashArrowArc(422,77)(6,-120,240){1}
\PhotonArc(410,50)(30,78,123){2}{3.5}
   \DashArrowArc(389,72)(6,-45,315){1}
\PhotonArc(410,50)(30,146,180){2}{3}
    \Line(410,79)(410,95)
      \Vertex(410,79){1}\Text(410,75)[t]{$\eps$}
\Line(380,50)(440,50)
\Photon(440,50)(460,50){2}{3}
  \Vertex(440,50){1}\Text(440,45)[t]{$\eps$}
\Text(410,20)[]{(i)}
\end{picture} 
\end{center}
\begin{center}
\begin{picture}(500,100)(50,-10)
\Photon(100,50)(120,50){2}{3}
  \Vertex(120,50){1}\Text(115,45)[rt]{$\eps n_F e_r^2$}
\CArc(140,36)(25,33,147)
    \Line(120,50)(120,67)
  \PhotonArc(140,50)(20,180,247){-2}{4}
     \DashArrowArc(140,30)(8,-90,270){1}
  \PhotonArc(140,50)(20,293,360){2}{4}
\Photon(160,50)(180,50){2}{3}
  \Vertex(160,50){1}\Text(165,45)[lt]{$n_F e_r^2$}
\Text(140,0)[]{(j)}

\Photon(210,50)(230,50){2}{3}
  \Vertex(230,50){1}\Text(225,45)[rt]{$\eps n_F e_r^2$}
\CArc(255,34)(30,31,149)
    \Line(230,50)(230,67)
  \PhotonArc(255,50)(25,180,212){-2}{2}
   \DashArrowArc(240,33)(7,-120,240){1}
  \PhotonArc(255,50)(25,246,294){2}{3} 
   \DashArrowArc(270,33)(7,-60,300){1}
  \PhotonArc(255,50)(25,328,360){2}{2}
\Photon(280,50)(300,50){2}{3}
  \Vertex(280,50){1}\Text(285,45)[lt]{$\eps$}
\Text(255,0)[]{(k)}

\Photon(330,50)(350,50){2}{3}
  \Vertex(350,50){1}\Text(345,45)[rt]{$\eps^2$}
\CArc(375,34)(30,31,149)
    \Line(350,50)(350,67)
  \PhotonArc(375,50)(25,180,212){-2}{2}
   \DashArrowArc(360,33)(7,-120,240){1}
  \PhotonArc(375,50)(25,246,294){2}{3} 
   \DashArrowArc(390,33)(7,-60,300){1}
  \PhotonArc(375,50)(25,328,360){2}{2}
\Photon(400,50)(420,50){2}{3}
  \Vertex(400,50){1}\Text(405,45)[lt]{$n_F e_r^2$}
\Text(375,0)[]{(l)}
\end{picture}
\end{center}
\begin{center}
\begin{picture}(300,100)(0,-10)
\Photon(50,50)(70,50){2}{3}
  \Vertex(70,50){1}\Text(65,45)[rt]{$\eps^2$}
\CArc(100,33)(35,30,150)
   \Line(70,50)(70,67)
  \PhotonArc(100,50)(30,180,209){-2}{2.5}
     \DashArrowArc(77,30)(6,-135,255){1}
  \PhotonArc(100,50)(30,231,259){2}{2.5} 
     \DashArrowArc(100,20)(6,-90,270){1}
  \PhotonArc(100,50)(30,281,309){2}{2.5}
     \DashArrowArc(123,30)(6,-45,315){1}
  \PhotonArc(100,50)(30,331,360){-2}{2.5}
\Photon(130,50)(150,50){2}{3}
  \Vertex(130,50){1}\Text(135,45)[lt]{$\eps$}
\Text(100,0)[]{(m)} 

\Photon(180,40)(220,40){2}{6}
\Line(200,40)(200,60)
\GCirc(200,40){4}{1}
  \Line(197,43)(203,37)
  \Line(197,37)(203,43)
\Text(200,0)[]{(n)}
\end{picture} \\
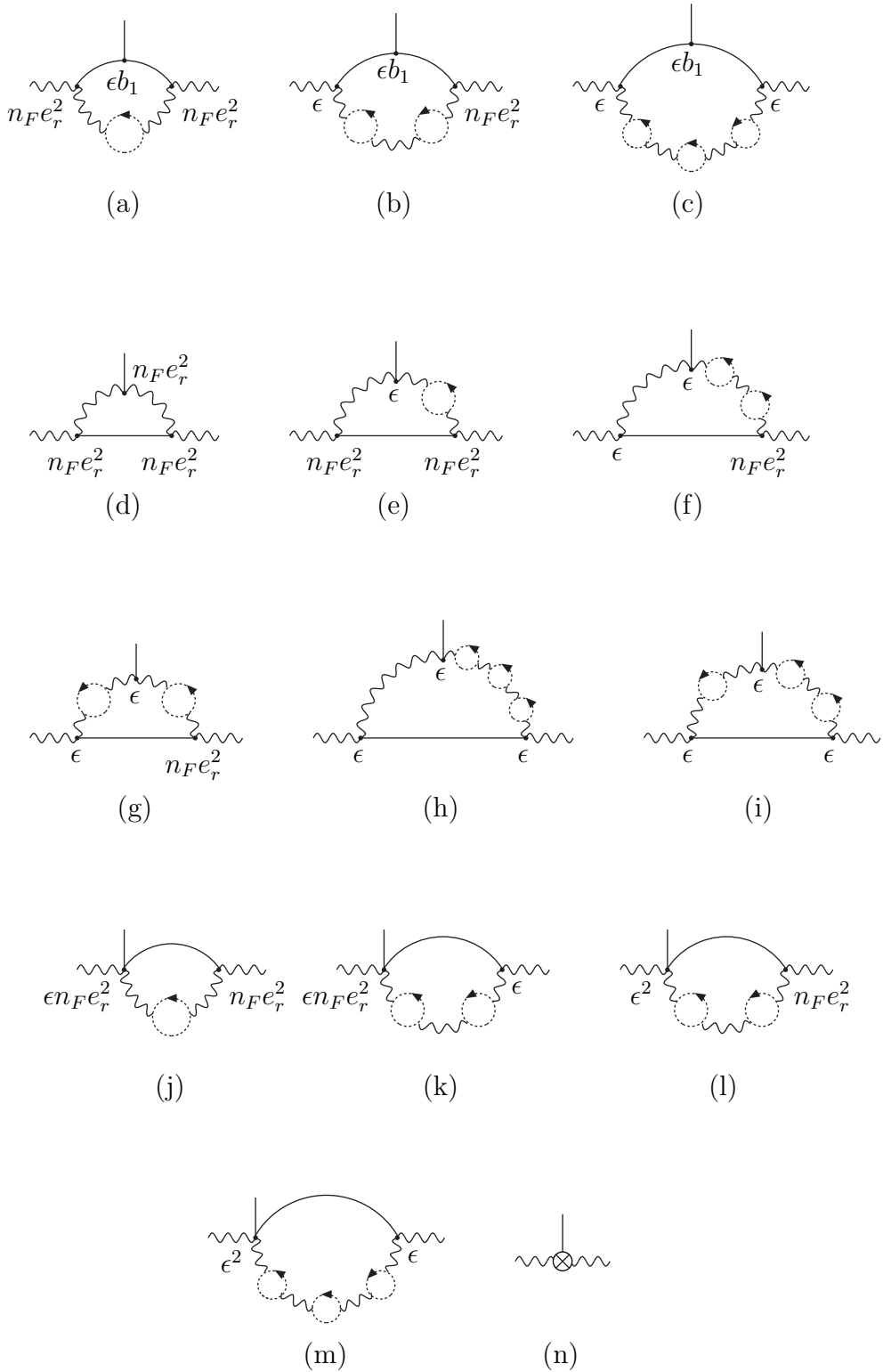
Fig. 12.  $O(e_r^6)$ $\phi \!-\!A\!-\!A$ vertex diagrams 
with internal conformal modes.  
\end{center}

From a direct calculation, we can see that the sum of 
the UV divergences from diagrams (d) to (i) in Fig. 12  
exactly vanishes when evaluated at zero momentum 
for the external conformal mode. 
The sum of the UV divergences from all other diagrams 
also vanishes:
\bb
   \Gamma^{\phi AA}_{\mu\nu}(0,k,-k) \vert_{II} 
    = \biggl\{ -\fr{8}{81}+\fr{16}{81}-\fr{8}{81} \biggr\}
       \fr{n_F^3}{b_c}\fr{e_r^6}{(4\pi)^6}\fr{1}{\eps} 
       \bigl( \dl_{\mu\nu} k^2 -k_{\mu}k_{\nu} \bigr) 
    =0.  
\ee
The first value in the braces results from the sum of (a) to (c),  
the second from the sum of (j) to (m), and  
the third from the direct counterterm diagram (n),   
whose value is derived from the residue of the double 
pole of $Z_3$.  

\section{Physical States in 4D Quantum Gravity} 
\setcounter{equation}{0}
\noindent

That a theory possesses quantum diffeomorphism invariance implies  
that the theory is independent of the background metric. 
Thus, in the quantum gravity phase, the physical distance measured by 
the metric field would become meaningless.  
This implies that IR effects and 
UV effects are sufficiently strong. 
Therefore, the usual Fock states based on 
the vanishing of correlations among them in the IR limit are 
no longer physical states in quantum gravity.  

The physical state conditions reflecting background-metric 
independence are represented by the equation 
$\fr{\dl Z}{\dl\hg^{\mu\nu}} = \langle {\hat T}_{\mu\nu} \rangle =0$. 
These are usually called the Hamiltonian momentum constraints.  
In 4D quantum gravity, we use perturbation theory to treat the traceless 
mode so that the background-metric independence is slightly violated. 
This violation would, in principle, vanish if the perturbative calculation 
were carried out to infinitely high order.  
However, as in 2D quantum gravity~\cite{dk,h93}, 
we can prove rigorously the background-metric independence 
for the conformal mode in the free 
field limit, $t_r=e_r=0$~\cite{amm1,hs}, 
which is realized at very high energies.  
This is an advantage of our formalism. 
We can study the structure of the physical states in the very high-energy 
regime using this free field model. 

The physical state conditions are exactly solved in 2D quantum gravity 
using conformal field theory~\cite{kpz,dk}. 
Recall that although 2D quantum gravity is a free 
theory, it has rich structure: an infinite number of physical 
states~\cite{bmp}, a non-linear property of factorization~\cite{s}, 
and so on.  For instance, the physical states of 2D quantum gravity 
coupled to a scalar field $X$ with central charge $c_X=1$ are 
given by 
  $\int d^2 x \!:\!\e^{2\phi} \!:$, 
  $\int d^2 x \!:\!\pd X {\bar \pd} X \!:$, 
   $\int d^2 x \!:\!\bigl( \pd^2 X \pm 2i (\pd X)^2 \bigr) 
      \bigl( {\bar \pd}^2 X \pm 2i ({\bar \pd} X)^2 \bigr) 
        \e^{-\phi \mp iX} \!:$,  
and so on, where $: :$ denotes the normal order, and  
the conformal mode is properly normalized. These infinitely numerous 
states are called ``discrete states." 
Here, note that there is an analogy 
between this two-dimensional field $X$ and the traceless mode 
$h^{\mu}_{~\nu}$ in 4D quantum gravity, because $2n$-th order 
fields in $2n$ dimensions have the same IR and UV behavior 
for every $n$. 
As for the conformal mode, the relation to 2D quantum gravity is 
more apparent: the first term in the action, $S_1(\phi,\bg)$, 
in each dimension satisfies the Wess-Zumino consistency 
condition, which is used to prove the background-metric 
independence for the conformal mode~\cite{h93,amm1,hs}. 
Thus, although 4D quantum gravity is asymptotically free, 
the physical states are non-trivial even in the very high-energy 
regime, which can be described with 
diffeomorphism invariant composite fields.  
  
One of the simplest physical states is the cosmological constant. 
We here investigate it for $t_r=e_r=0$ in detail.  
In dimensional regularization, the cosmological constant is given 
by $\Lam \int d^D x \e^{D\phi}
=\Lam \sum _n \int d^D x \fr{D^n}{n!} \phi^n$  
in the flat background,\footnote{ 
In previous papers~\cite{am,h00}, 
in analogy with the operator formalism of DDK,  
the anomalous dimension was computed using a procedure in which 
the deformed cosmological constant field $\e^{\a \phi}$ is considered, 
and the equation $\a=4+\gm_{\Lam}$ is solved  
for $\a$, where $\gm_{\Lam}$ is given by a function of $\a$ 
computed with one-loop diagrams. 
However, this procedure is actually incorrect in dimensional 
regularization, because diffeomorphism invariance  is obviously 
violated, though the derived anomalous dimension 
for the cosmological constant seems correct. 
Here, we revise the procedure in a manifestly invariant way.   
} 
which is renormalized by replacing the bare 
constant $\Lam$ by $Z_{\Lam} \Lam_r$. 
Since there is no coupling constant for the conformal mode,  
we, for the present, evaluate the renormalization factor $Z_{\Lam}$ 
for large $n_F$. 

The diagrams yielding simple poles at $O(1/b_c)$ 
and $O(1/b_c^2)$ are given by (a) and (b) 
in Fig. 13, respectively. 
Contrastingly, diagrams with separated ovals, for example,  
diagrams (a) and (b), in Fig. 14 at $O(1/b_c^2)$, 
do not produce simple poles. More precisely, 
the sum of the UV divergences from diagram (a), itself, in Fig. 14  
and from the associated subdivergence counterterm diagram 
not depicted here, yields only a double pole, 
while diagram (b) in Fig. 14 gives no contribution to $Z_{\Lam}$, 
because the UV divergence from diagram (b) itself cancels that 
from its associated subdivergence counterterm diagram.  
In general, because a diagram with more than two separated ovals does not 
produce a simple pole, it does not contribute to the 
anomalous dimension.  Thus, we obtain   
\bb
   Z_{\Lam} =1
     - \fr{2}{\tb_c} \fr{1}{\eps} -\fr{2}{\tb_c^2} \fr{1}{\eps} 
     + \fr{2}{\tb_c^2}\fr{1}{\eps^2} +\cdots,  
\ee
and hence the anomalous dimension of the cosmological constant is 
\bb
   \gm_{\Lam} = \fr{\mu}{Z_{\Lam}}\fr{d Z_{\Lam}}{d \mu}
              = \fr{4}{b_c} + \fr{8}{b_c^2} + \cdots.
\ee 
\begin{center}
\begin{picture}(200,110)(0,-20)
\CArc(50,50)(20,0,360)
\Vertex(50,30){1}
\Text(52,12)[]{$\cdots$}
\Line(38,10)(50,30)
\Line(42,9)(50,30)
\Line(62,10)(50,30)
\Text(50,-10)[]{(a)}

\CArc(150,50)(20,0,360)
\Line(150,30)(150,70) 
   \Vertex(150,70){1}\Text(150,75)[b]{$\eps b_1$}
\Vertex(150,30){1}
\Text(152,12)[]{$\cdots$}
\Line(138,10)(150,30)
\Line(142,9)(150,30)
\Line(162,10)(150,30)
\Text(150,-10)[]{(b)}
\end{picture} \\
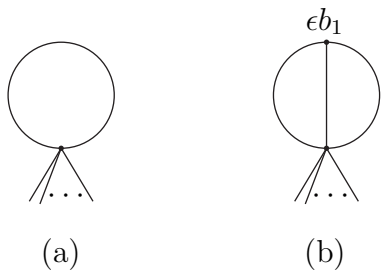
Fig. 13. $O(1/b_c)$ and $O(1/b_c^2)$ corrections 
to the cosmological constant.
\end{center}
\begin{center}
\begin{picture}(300,100)(0,-10)
\Oval(83,50)(22,12)(40)
\Oval(117,50)(22,12)(-40)
\Vertex(100,38){1}
\Text(102,12)[]{$\cdots$}
\Line(88,10)(100,38)
\Line(92,9)(100,38)
\Line(112,10)(100,38)
\Text(100,-3)[]{(a)}

\Oval(183,50)(22,12)(40)
\Oval(217,50)(22,12)(-40)
\Line(228,68)(228,85) 
    \Vertex(228,68){1}\Text(232,72)[l]{$\eps b_1$}
\Vertex(200,38){1}
\Text(202,12)[]{$\cdots$}
\Line(188,10)(200,38)
\Line(192,9)(200,38)
\Line(212,10)(200,38)
\Text(200,-3)[]{(b)}
\end{picture} \\ 
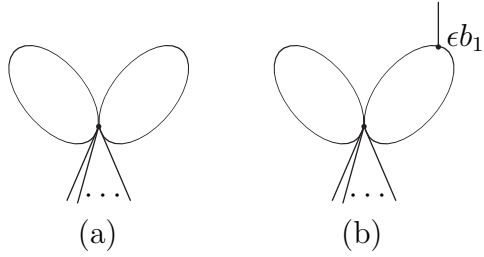
Fig. 14. $O(1/b_c^2)$ corrections to the cosmological 
constant that do not yield simple poles. 
\end{center} 

Let us now see what is obtained when summing up all orders. 
Using the analogy to 2D quantum gravity, the anomalous dimension 
defined by subtracting the canonical dimension from the quantum dimension  
was conjectured to be~\cite{am,h00}~\footnote{ 
In two dimensions, the anomalous dimension can be exactly calculated 
using conformal field theory as $\gm_{\Lam}=\a-2$, where 
$\a=b_c (1-\sq{1-4/b_c})$ is the Liouville charge of the 
cosmological constant operator, $:\! \e^{\a\phi}\!:$~\cite{kpz,dk}. 
For 2D quantum gravity coupled to $N$ scalar fields, $b_c=(25-N)/6$. 
In dimensional regularization, the first two terms of $\gm_{\Lam}$ 
in the expansion w.r.t. $1/b_c$ is also obtained by evaluating 
the same diagrams in Fig. 13 when using the propagator and the 
vertices obtained from the bare action, 
$\fr{b_1}{D-2}\int d^D x \sq{g}R$ and the series (\ref{series(2)}), 
where $b_1=\fr{b_c}{4\pi}$ can be determined by computing one-loop 
divergences for the counterterm of $\sq{\hg}\hR$.     
} 
\bba
   \gm_{\Lam} &=& 2b_c \biggl( 1-\sq{1-\fr{4}{b_c}} \biggr) -4
                    \nonumber \\ 
       &=& \fr{4}{b_c}+\fr{8}{b_c^2}+\fr{20}{b_c^3}+\cdots.
\eea
The first two terms in the expansion agree exactly with the 
computed terms. 
The corresponding diagrams at $O(1/b_c^m)$ are given by 
diagrams with a single oval in Fig. 15. 
Here, the first diagram has $m-1$ 
$\phi^3$-vertices with $\eps b_1$,    
the second one has $m-3$ $\phi^3$ vertices with $\eps b_1$ and   
a $\phi^4$-vertex with $\eps^2 b_1$, and so on. The last one is 
constructed  by using a $\phi^{m+1}$-vertex with $\eps^{m-1} b_1$. 
Each diagram has $m$ loops multiplied by $\eps^{m-1}$, so that 
it yields a simple pole. 
Since the other potentially divergent diagrams with multiple ovals 
do not produce simple poles, only the diagrams in Fig. 15  
contribute to the anomalous dimension at $O(1/b_c^m)$.    
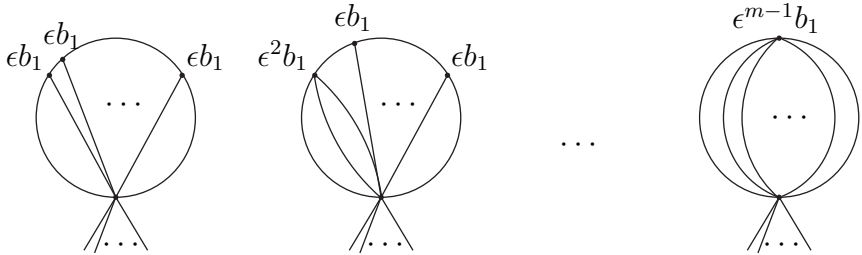
\begin{center}
\begin{picture}(350,120)(0,-10)
\CArc(50,60)(30,0,360)
  \Line(50,30)(25,76)
      \Vertex(25,76){1}\Text(23,80)[rb]{$\eps b_1$}
  \Line(50,30)(30,82)
      \Vertex(30,82){1}\Text(30,88)[b]{$\eps b_1$}
  \Text(53,65)[]{$\cdots$}
  \Line(50,30)(75,76) 
      \Vertex(75,76){1}\Text(77,80)[lb]{$\eps b_1$}
\Vertex(50,30){1}
\Text(52,12)[]{$\cdots$}
\Line(38,10)(50,30)
\Line(42,9)(50,30)
\Line(62,10)(50,30)

\CArc(150,60)(30,0,360)
  \CArc(200,87)(76,188,228)
  \CArc(75,19)(76,8,49) 
     \Vertex(125,76){1}\Text(123,80)[rb]{$\eps^2 b_1$}
  \Line(150,30)(140,88) 
     \Vertex(140,88){1}\Text(140,95)[b]{$\eps b_1$}
  \Text(157,65)[]{$\cdots$}
  \Line(150,30)(175,76) 
     \Vertex(175,76){1}\Text(177,80)[lb]{$\eps b_1$}
\Vertex(150,30){1}
\Text(152,12)[]{$\cdots$}
\Line(138,10)(150,30)
\Line(142,9)(150,30)
\Line(162,10)(150,30)

\Text(225,50)[]{$\cdots$}

\CArc(300,60)(30,0,360)
 \CArc(311,60)(32,112,248)
 \CArc(326,60)(40,131,229)
 \Text(305,60)[]{$\cdots$}
 \CArc(289,60)(32,-68,68)
    \Vertex(300,90){1}\Text(300,94)[b]{$\eps^{m-1}b_1$}
\Vertex(300,30){1}
\Text(302,12)[]{$\cdots$}
\Line(288,10)(300,30)
\Line(292,9)(300,30)
\Line(312,10)(300,30)
\end{picture} \\
Fig. 15. $O(1/b_c^m)$ corrections to the cosmological constant.
\end{center}

Here, there is an important observation concerning the ambiguity of 
the action $G_D$. 
This action is defined by the series (\ref{series(4)}). 
All vertices used in Fig. 15 stem from the term 
$\phi^n \bDelta_4 \phi$ in $S_n^{(4)}$. 
{}From a simple counting of the number of loops and $\eps$'s 
at the vertices, we can easily see that, if we use the vertices 
coming from the 
$O(\phi^n)$ term in $S_n^{(4)}$, diagrams with a single oval 
become finite. Also, if we use vertices obtained 
when $b_0 \neq 0$ in (\ref{b-series}), 
the same conclusion is reached.  
Furthermore,  although there is a potentially divergent diagram 
with the $O(\phi^n)$ vertices in the case that it has more than two 
separated ovals, it does not yield a simple pole, as mentioned above. 
Thus, the anomalous dimensions of the cosmological constant 
can be determined only from the first term in $S_n^{(4)}$. 
 
In this paper, although we have only considered the case of the cosmological 
constant, generalization is straightforward. 
It seems natural to assume that the physical state is given by 
a combination of diffeomorphism invariant composite fields. 
Thus, we conjecture that 
the physical state can be classified according to the condition of 
renormalizability. In that case,  
all physical quantities, such as anomalous dimensions, would be 
independent of the choice of the $O(\phi^n)$ term in $S_n^{(4)}$, 
as in the case of the cosmological constant. 
In this sense, $G_D$ is almost unique.

\section{Discussion} 
\setcounter{equation}{0}
\noindent

In this paper, we studied the problems of higher-order 
renormalization and physical states in 4D quantum gravity.   
Using dimensional regularization with great care concerning 
the conformal-mode dependence, 
we carried out the renormalization of a model of 4D quantum gravity 
coupled to QED up to $O(e_r^6)$ and $O(t_r^2)$. 
The procedure is manifestly diffeomorphism invariant, 
and resummation was done without making an DDK-like assumption. 
As a by-product, we found that higher-order gravitational corrections 
to the beta functions of $e_r$ and $t_r$ are negative. 

This model suggests that a configuration characterized by the vanishing of 
the Weyl tensor becomes dominant at very high energies, where the 
coupling $t_r$ is small. Such a configuration is called  conformally 
flat, or $S^4 \times ({\rm conformal ~factor})$. 
Quantum gravity may be realized in the early universe and 
on inside of black holes, where spacetime singularities would appear 
in the classical description. 
However, this model suggests that such  singularities 
disappear through the effect of quantum gravity~\cite{h01b}.   

Physical states in our model were discussed  
in analogy with those in 2D quantum gravity. 
In particular, the cosmological constant term was studied in detail. 
In general, even though the theory is asymptotically free, 
the physical states might be given by diffeomorphism-invariant 
composite fields, such as glueball states in 4D Yang-Mills 
theory, because of the strong IR effects.   
Therefore, to distinguish them from ordinary gravitons, we here call 
them ``graviball states." 
A photon may also be confined as a photonball state in quantum 
gravity. 
On the other hand, usual graviton and photon states are low-energy 
states, which would appear when the background-metric independence, 
namely the condition ${\hat T}_{\mu\nu} |phys \rangle =0$, is violated. 
It is not, at present, understood how to connect the quantum gravity phase 
with our present Einstein phase. 
It seems that there is a discontinuity between them.   

Finally, we comment on recent interesting numerical results  
concerning the dynamical triangulation (DT) approach~\cite{wei,bk,adj} 
in four dimensions~\cite{migdal}--\cite{hagura}. 
It is believed that the DT method realizes the quantum gravity phase 
in the very high-energy regime. 
The entropy factor in the DT approach corresponds to the four-derivative 
gravitational actions, because they are dimensionless in four dimensions, 
so that they essentially represent quantum effects.  
The annomalous dimension of the 
cosmological constant in this regime is almost given by a function of 
$b_c$, as discussed in Section 7, and corrections that depend on 
the coupling constant $t_r$ are small~\cite{h00}. 
If we consider a gravity system 
coupled to $n_X$ conformal scalar fields and $n_A$ gauge fields, 
$b_c$ is given by a function of $n_X +62 n_A$, and hence 
$\gm_{\Lam}$ is a function of $n_X +62 n_A$. 
This number dependence should be numerically checked using  
the DT approach~\cite{hey}.

\vspace{5mm}

\begin{flushleft}
{\bf Acknowledgements}
\end{flushleft}

This work was supported in part by a Grant-in-Aid 
for Scientific Research from the Ministry of Education, Science 
and Culture of Japan. 
  
\begin{center}
{\Large {\bf Appendix}}
\end{center}

\appendix 
\section{Conformal-mode Dependence of Curvatures}
\setcounter{equation}{0}
\noindent

The conformal mode dependence of curvatures in $D$ dimensions is 
obtained by using the expression for the Christoffel symbol  
$\Gamma^{\lam}_{~\mu\nu} 
    = {\bar \Gamma}^{\lam}_{~\mu\nu}
       +\bg^{\lam}_{~\mu}\bnabla_{\nu}\phi 
       +\bg^{\lam}_{~\nu}\bnabla_{\mu}\phi
       -\bg_{\mu\nu} \bnabla^{\lam}\phi $, 
as 
\bba
     R^{\lam}_{~\mu\s\nu}
   &=& \bR^{\lam}_{~\mu\s\nu} + \bg^{\lam}_{~\nu}\bDelta_{\mu\s}   
       -\bg^{\lam}_{~\s}\bDelta_{\mu\nu}+\bg_{\mu\s}\bDelta^{\lam}_{~\nu}
       -\bg_{\mu\nu}\bDelta^{\lam}_{~\s} 
               \nonumber  \\
   && \qquad\qquad\qquad 
      + \bigl( \bg^{\lam}_{~\nu}\bg_{\mu\s}
            -\bg^{\lam}_{~\s}\bg_{\mu\nu} \bigr) 
               \bnabla_{\dl}\phi\bnabla^{\dl}\phi, 
                   \\ 
    R_{\mu\nu} 
    &=&\bR_{\mu\nu}-(D-2)\bDelta_{\mu\nu} 
        -\bg_{\mu\nu} \bigl\{ \bBox \!\!\phi 
        +(D-2)\bnabla_{\lam}\phi \bnabla^{\lam}\phi \bigr\},
                       \\ 
   R &=& \e^{-2\phi} \bigl\{ \bR -2(D-1)\bBox \!\!\phi 
        -(D-1)(D-2)\bnabla_{\lam}\phi \bnabla^{\lam}\phi \bigr\}, 
                  \label{R}
\eea 
where $\bDelta_{\mu\nu}=\bnabla_{\mu}\bnabla_{\nu}\phi 
-\bnabla_{\mu}\phi\bnabla_{\nu}\phi$. 
Our curvature conventions are $R_{\mu\nu}=R^{\lam}_{~\mu\lam\nu}$ 
and $R^{\lam}_{~\mu\s\nu}=\pd_{\s}\Gamma^{\lam}_{~\mu\nu}-\cdots$.

The variation formulae with respect to the conformal change 
$\dl_{\om}g_{\mu\nu} =2\om g_{\mu\nu}$ are given by 
\bb
  \dl_{\om} \sq{g}R = (D-2)\om \sq{g}R -2(D-1)\sq{g}\Box \om   
\ee
and 
\bba 
   \dl_{\om} \sq{g}R^{\mu\nu\lam\s}R_{\mu\nu\lam\s} 
     &=& (D-4) \om \sq{g}R^{\mu\nu\lam\s}R_{\mu\nu\lam\s} 
      -8\sq{g} R^{\mu\nu}\nabla_{\mu}\nabla_{\nu} \om , 
           \\ 
   \dl_{\om} \sq{g}R^{\mu\nu}R_{\mu\nu} 
     &=& (D-4) \om \sq{g}R^{\mu\nu}R_{\mu\nu} 
         -2 \sq{g}R \Box \om 
           \nonumber  \\ 
     && \qquad\qquad\qquad
      -2(D-2)\sq{g} R^{\mu\nu}\nabla_{\mu}\nabla_{\nu} \om ,    
            \\ 
   \dl_{\om} \sq{g}R^2
     &=& (D-4) \om \sq{g}R^2 - 4(D-1) \sq{g}R \Box \om , 
            \\ 
   \dl_{\om} \sq{g}\Box \!R 
     &=& (D-4)\om \sq{g} \Box \!R 
         +(D-6)\sq{g} \nabla^{\lam}R \nabla_{\lam}\om 
             \nonumber \\ 
  && \qquad\qquad\qquad
         -2 \sq{g}R \Box \om -2(D-1) \sq{g}\Box^2 \!\om , 
             \\ 
  \dl_{\om} \sq{g}F_{\mu\nu}F^{\mu\nu} 
      &=& (D-4) \om \sq{g}F_{\mu\nu}F^{\mu\nu}.
\eea 
Applying these variation formulae to (\ref{dl-gm}), 
we obtain 
\bba
   [\dl_{\om_1}, \dl_{\om_2}] \Gm 
    &=& \bigl\{ 4 \eta_1 +D \eta_2 +4(D-1)\eta_3 +(D-4)\eta_4 \bigr\} 
            \nonumber  \\ 
    && \qquad \times 
     \int d^D x \sq{g} R ( \om_1 \Box \om_2 -\om_2 \Box \om_1). 
\eea
Thus, the integrability condition is given by (\ref{integrability}).

\section{Expansions of Curvatures}
\setcounter{equation}{0}
\noindent

{}From expression (\ref{R}), we can see that   
the spacetime integral of the scalar curvature is expanded around $D=2$  
as follows:   
\bb
   \int d^D x \sq{g}R 
   = \sum_{n=0}^{\infty} \fr{(D-2)^2}{n!}
       \int d^D x \sq{\hg} \Bigl\{ 
         -(D-1)\phi^n \bBox \! \phi + \bR \phi ^n  \Bigr\}. 
\ee
This is the action of 2D quantum gravity near two dimensions. 

Next, we consider the expansion of four-derivative actions 
near four dimensions.  
The spacetime integral of $M_D$ is expanded around $D=4$ as  
\bba
   && \int d^D x \sq{g} M_D 
             \nonumber \\
   && = -\fr{D-4}{4(D-1)} \int d^D x \sq{g} R^2 
             \nonumber   \\ 
   && = -\fr{1}{4(D-1)} \sum_{n=0}^{\infty} \fr{(D-4)^n}{n!}
        \int d^D x \sq{\hg} \biggl\{ 
        (D-4)\phi^n \bR^2 -2(D-1)(D-6)\phi^n \bR\bBox \!\phi 
             \nonumber \\ 
   && \qquad\qquad\qquad\qquad 
        +2(D-1)(D-2)\phi^n \bnabla^{\lam}\bR \bnabla_{\lam} \phi 
        +4(D-1)^2 \phi^n \bBox^2 \!\phi 
             \nonumber  \\ 
   && \qquad\qquad\qquad\qquad
        +8(D-1)^2(D-4) \phi^n \bBox \! \phi 
               \bnabla^{\lam}\phi \bnabla_{\lam}\phi 
             \nonumber  \\
   && \qquad\qquad\qquad\qquad 
        +(D-1)^2(D-2)^2(D-4) \phi^n 
              (\bnabla^{\lam}\phi \bnabla_{\lam}\phi )^2 
          \biggr\}, 
            \label{series-MD}
\eea
and, also, the spacetime integral of $G_4$ is expanded around $D=4$ as  
\bba
   && \int d^D x \sq{g} G_4 
             \nonumber \\
   && = \sum_{n=0}^{\infty} \fr{(D-4)^n}{n!}
        \int d^D x \sq{\hg} \biggl\{ 
          \phi^n \bG_4 
          +4(D-3)\phi^n \bR^{\mu\nu}\bnabla_{\mu}\bnabla_{\nu} \phi 
             \nonumber \\ 
   && \qquad\qquad\qquad 
        -2(D-3)\phi^n \bR\bBox \!\phi 
        -2(D-2)(D-3)(D-4) \phi^n \bBox \! \phi 
               \bnabla^{\lam}\phi \bnabla_{\lam}\phi 
             \nonumber  \\
   && \qquad\qquad\qquad 
        -(D-2)(D-3)^2(D-4)\phi^n 
              (\bnabla^{\lam}\phi \bnabla_{\lam}\phi )^2 
          \biggr\}. 
          \label{series-G4}
\eea
The last two terms in each of the series in (\ref{series-MD}) 
and (\ref{series-G4}) give contributions of $O(\phi^{n+2})$ 
and $O(\phi^{n+3})$ 
at order $(D-4)^n$, while the series in (\ref{S(4)}), which we seek, 
gives a contribution of at most $O(\phi^{n+1})$. 
Thus, in order to obtain the series in (\ref{S(4)}), we must consider 
the sum of these two actions, $\int d^D x \sq{g} (G_4+\eta M_D)$,  
and choose the constant $\eta$ properly in order to remove 
the last two terms.  
Although we cannot exactly remove these terms simultaneously,  
the series we seek can be obtained by merely removing the last term, 
so that $\eta=-\fr{4(D-3)^2}{(D-1)(D-2)}$.  
In this way, we obtain     
\bba
  && \int d^D x \sq{g}G_D 
           \nonumber \\ 
  && = \sum_{n=0}^{\infty} \fr{(D-4)^n}{n!} 
       \int d^D x \sq{\hg} 
        \biggl\{ \phi^n \bE_D 
        +\fr{4(D-3)^2}{D-2}\phi^n \bBox^2 \!\!\phi 
        +4(D-3)\phi^n \bR^{\mu\nu}\bnabla_{\mu}\bnabla_{\nu}\phi 
            \nonumber  \\ 
  && \qquad\qquad
       -\fr{4(D-3)(D^2-6D+10)}{(D-1)(D-2)}\phi^n \bR\bBox \!\!\phi 
      -\fr{2(D-3)^2(D-6)}{(D-1)(D-2)}
           \phi^n \bnabla^{\lam}\bR \bnabla_{\lam}\phi 
            \nonumber  \\ 
  && \qquad\qquad\qquad
         -\fr{2(D-3)(D-4)^3}{D-2} \phi^n \bBox \!\!\phi
            \bnabla_{\lam}\phi\bnabla^{\lam}\phi   
     \biggr\} 
                \nonumber \\ 
  && = \int d^D x \sq{\hg} \biggl\{ 
         \bG_4 + (D-4) \biggl( 2\phi\bDelta_4 \phi +\bE_4 \phi 
                               + \fr{1}{18}\bR^2 \biggr) 
              \label{GD-B} \\ 
  && \qquad\qquad\qquad 
        +\half(D-4)^2 \biggl( 2\phi^2 \bDelta_4 \phi + \bE_4 \phi^2  
              \nonumber  \\ 
  && \qquad\qquad\qquad\qquad
         + 6\phi \bBox^2 \!\!\phi  
         +8 \phi \bR^{\mu\nu}\bnabla_{\mu}\bnabla_{\nu}\phi  
        -\fr{28}{9} \phi\bR \!\bBox \! \phi 
        +\fr{8}{9} \phi \bnabla^{\lam}\bR \bnabla_{\lam}\phi 
              \nonumber  \\ 
   && \qquad\qquad\qquad\qquad\qquad 
     -\fr{14}{9} \bR \bBox \!\phi  +\fr{1}{9} \bR^2 \phi 
       +\fr{5}{54}\bR^2 \biggr)  
              \nonumber \\ 
   && \qquad\qquad\qquad   
       + \fr{1}{3!}(D-4)^3 \Bigl( 
       2\phi^3 \bDelta_4 \phi + \bE_4 \phi^3 
       -6 \bBox \! \phi \bnabla^{\lam}\phi \bnabla_{\lam}\phi 
       + 9 \phi^2 \bBox^2 \!\!\phi +\cdots \Bigr) 
              \nonumber  \\ 
   && \qquad\qquad\qquad 
       + O((D-4)^4)  \biggr\}. 
              \nonumber 
\eea

\section{BRST Formulation}
\setcounter{equation}{0}
\noindent 

The general coordinate transformations 
$\dl_{\xi} g_{\mu\nu}=g_{\mu\lam}\nabla_{\nu}\xi^{\lam} 
+ g_{\nu\lam}\nabla_{\mu}\xi^{\lam}$  
can be completely decomposed into transformations of the conformal 
mode and the traceless mode as  
\bba
   \dl_{\xi} \bg_{\mu\nu} &=& \bg_{\mu\lam}\bnabla_{\nu}\xi^{\lam} 
                 +\bg_{\nu\lam}\bnabla_{\mu}\xi^{\lam}  
                 -\frac{2}{D}\bg_{\mu\nu}\hnabla_{\lam}\xi^{\lam} ,    
                    \nonumber     \\ 
   \dl_{\xi} \phi &=&  \xi^{\lam}\pd_{\lam}\phi  
                       + \frac{1}{D} \hnabla_{\lam}\xi^{\lam}, 
                      \label{gct}
\eea  
where $\bnabla_{\lam}\xi^{\lam}=\hnabla_{\lam}\xi^{\lam}$ is used. 
The traceless-mode part can be expanded in  
the coupling constant $t$. 
The general coordinate transformation for $A_{\mu}$ is given by 
\bb
   \dl_{\xi} A_{\mu} = \xi^{\nu}\nabla_{\nu}A_{\mu} 
                       + A_{\nu} \nabla_{\mu} \xi^{\nu}
                     = \xi^{\nu}\pd_{\nu}A_{\mu} 
                       + A_{\nu} \pd_{\mu} \xi^{\nu}, 
                  \label{gct-A}
\ee
and the $U(1)$ gauge transformation is 
$\dl_{\lam} A_{\mu} = \pd_{\mu}\lam$. 

Let us fix the coordinate transformation invariance for the traceless 
mode and  the gauge invariance of the photon field 
according to the procedure of the BRST quantization. 
The BRST transformations can be obtained by replacing $\xi^{\mu}/t$ 
in the equations for general coordinate transformations, (\ref{gct}) 
and (\ref{gct-A}), and $\lam$ in the $U(1)$ gauge transformation 
with corresponding Grassmann ghost fields, $c^{\mu}$ and $c$, 
respectively. We obtain    
\bba
    {\bf \dl_B} h^{\mu}_{~\nu}  
       &=&  \biggl\{ \hnabla^{\mu} c_{\nu} 
                       +\hnabla_{\nu} c^{\mu}
           - \fr{2}{D} \dl^{\mu}_{~\nu}  
                    \hnabla_{\lam} c^{\lam}   
           + t c^{\lam} \hnabla_{\lam} h^{\mu}_{~\nu}   
                \nonumber    \\ 
    && \quad 
           + \frac{t}{2} h^{\mu}_{~\lam} 
                \Bigl( \hnabla_{\nu} c^{\lam} 
                  - \hnabla^{\lam} c_{\nu} \Bigr) 
           + \frac{t}{2} h^{\lam}_{~\nu} 
                \Bigl( \hnabla^{\mu} c_{\lam} 
                  - \hnabla_{\lam} c^{\mu} \Bigr) 
           + \cdots \biggr\},
               \nonumber   \\ 
     {\bf \dl_B} \phi 
       &=&  t c^{\mu} \pd_{\mu} \phi 
         + \frac{t}{D} \hnabla_{\mu} c^{\mu}, 
            \nonumber   \\ 
    {\bf \dl_B} A_{\mu} 
       &=& \pd_{\mu} c + t c^{\nu}\pd_{\nu} A_{\mu} 
                + t A_{\nu}\pd_{\mu}c^{\nu},       
          \label{brst} \\
    {\bf \dl_B} c^{\mu} 
       &=&t c^{\nu}\hnabla_{\nu} c^{\mu} 
          = t c^{\nu}\pd_{\nu} c^{\mu},   
            \nonumber   \\
    {\bf \dl_B} c 
       &=& t c^{\mu}\pd_{\mu} c,    
            \nonumber  \\
    {\bf \dl_B} {\tilde c}^{\mu} 
       &=&   B^{\mu},  
           \qquad   {\bf \dl_B} B^{\mu} = 0, 
              \nonumber     \\
    {\bf \dl_B} {\tilde c} 
       &=&   B,  
           \qquad   {\bf \dl_B} B = 0, 
              \nonumber  
\eea
where $c_{\mu}=\hg_{\mu\nu}c^{\nu}$. 
We can easily see that this BRST transformation is nilpotent. 
Using this transformation, the gauge-fixing term and the FP ghost 
action can be written as   
\bba
      I_{GF+FP}  
      &=&  \int d^D x \sq{\hg} ~{\bf \dl_B}
           \biggl\{ {\tilde c}^{\mu}{\hat N}_{\mu\nu} 
             \biggl(  \chi^{\nu} 
                 - \fr{\zeta}{2} B^{\nu} \biggr) 
              +  {\tilde c} \biggl( \hnabla^{\mu}A_{\mu} 
                    - \fr{\a}{2}B \biggr)  
                 \biggr\}
             \nonumber  \\ 
      &=& \int d^D x \sq{\hg} \biggl\{
          B^{\mu} {\hat N}_{\mu\nu} \chi^{\nu}   
            -\fr{\zeta}{2} B^{\mu} {\hat N}_{\mu\nu} B^{\nu}    
            - {\tilde c}^{\mu} {\hat N}_{\mu\nu} \hnabla^{\lam} 
              {\bf \dl_B} h^{\nu}_{~\lam}  \\ 
                          \label{gfix}
     && \qquad\qquad\qquad 
             + B \hnabla^{\mu}A_{\mu} - \fr{\a}{2} B^2 
             - {\tilde c} \hnabla^{\mu} {\bf \dl_B} A_{\mu}  
               \biggr\}, 
             \nonumber 
\eea
where $\chi^{\nu}=\hnabla^{\lam}h^{\nu}_{~\lam}$ and   
${\hat N}_{\mu\nu}$ is a symmetric second-order operator 
defined on $\hg_{\mu\nu}$. This action has a covariant form 
with respect to $\hg_{\mu\nu}$, but not with respect to the dynamical 
gravitational modes, $\phi$ and $h^{\mu}_{~\nu}$.  
Note that the conformal mode does 
not couple with all ghost fields, and 
the ghosts for the $U(1)$ gauge invariance are not coupled to the 
traceless mode either.  
If $B_{\mu}$ and $B$ are integrated out, we obtain~\cite{ft82} 
\bb
      I_{GF} = \int d^D x \sq{\hg} \biggl\{ 
         \fr{1}{2\zeta}  \chi^{\mu} {\hat N}_{\mu\nu} \chi^{\nu} 
         +\fr{1}{2\a} \Bigl( \hnabla^{\mu}A_{\mu} \Bigr)^2 \biggr\}.  
       \label{gfix}
\ee

In the following, we use the flat background, in which  
$\hg_{\mu\nu}=\dl_{\mu\nu}$. The bilinear term of the traceless mode 
is given by 
\bb
     \fr{1}{t^2}\int d^D x \bF_D = 
          \fr{D-3}{D-2} \biggl( h_{\mu\nu} \pd^4 h_{\mu\nu} 
         + 2\chi_{\mu} \pd^2 \chi_{\mu} 
         - \frac{D-2}{D-1} \chi_{\mu} \pd_{\mu}\pd_{\nu}\chi_{\nu}  
           \biggr) + O(t),
                  \label{kin}
\ee
where $\chi_{\mu}=\pd_{\lam}h_{\mu\lam}$ and  
$\pd^2 =\pd_{\lam}\pd_{\lam}$.  
{}From this expression, we should choose the following form 
for the gauge-fixing term: 
\bb
      {\hat N}_{\mu\nu}=  \fr{2(D-3)}{D-2} \biggl( 
                 -2 \pd^2\dl_{\mu\nu} 
           + \frac{D-2}{D-1} \pd_{\mu}\pd_{\nu} \biggr). 
\ee 
{}From expression (\ref{gfix}), $\zeta= 1$ corresponds to 
the Feynman-type gauge, in which the gauge-fixing term cancels out 
the last two of the kinetic terms in (\ref{kin}). 
Then, the propagator of the traceless mode becomes 
\bb
        \fr{D-2}{2(D-3)} \frac{1}{p^4} (I_H)_{\mu\nu, \lam\s}, 
\ee
where $I_H$ is the projection operator to the traceless mode, 
\bb
     (I_H)_{\mu\nu, \lam\s} 
        = \half \dl_{\mu\lam}\dl_{\nu\s} 
          + \half \dl_{\mu\s}\dl_{\nu\lam}
             - \frac{1}{D}\dl_{\mu\nu}\dl_{\lam\s},  
                  \label{IH}
\ee
which satisfies $I_H^2 =I_H$.

\section{Vertices in $b_1 S_1^{(4)}(\phi,\bg)$}
\setcounter{equation}{0}
\noindent
 
{}From the expansion in (\ref{GD-B}), we find   
\bb
     S_1^{(4)}(\phi,\bg) = \int d^D x \sq{\hg} 
      \biggl\{ 2 \phi \bDelta_4 \phi +\bE_4 \phi 
                      +\fr{1}{18}\bR^2 \biggr\}. 
\ee 
The bare action $b_1 S_1^{(4)}(\phi,\bg)$ is expanded in 
the coupling $t$ in the flat background, apart from the kinetic 
term of $\phi$, as 
\bba
    && {\cal L}_{S_1}^{2} = -\frac{2}{3} b_1 t 
           \pd^2 \phi \pd_{\mu}\pd_{\nu} h_{\mu\nu} 
           +\fr{b_1}{18} t^2 
           \bigl(\pd_{\mu}\pd_{\nu}h_{\mu\nu}\bigr)^2,
                    \\ 
    && {\cal L}_{S_1}^{3} = 2 b_1 t \biggl( 
          2\pd_{\mu} \phi \pd_{\nu} \pd^2 \phi 
          + \frac{4}{3}\pd_{\mu}\pd_{\lam}\phi \pd_{\nu}\pd_{\lam}\phi 
                  \nonumber \\ 
    && \qquad\qquad\qquad 
        -\frac{2}{3}\pd_{\lam}\phi \pd_{\mu}\pd_{\nu}\pd_{\lam} \phi 
          -2\pd_{\mu}\pd_{\nu}\phi \pd^2 \phi 
             \biggr)~ h_{\mu\nu} ,                
                    \\  
    && {\cal L}_{S_1}^{4} = 2 b_1 t^2 \Bigl( 
         \pd^2\phi\pd_{\mu}\pd_{\nu} \phi h_{\mu\lam} h_{\nu\lam} 
         +\pd_{\mu}\pd_{\nu}\phi \pd_{\lam}\pd_{\s}\phi 
               h_{\mu\nu}h_{\lam\s} 
               \nonumber \\ 
    && \qquad\qquad\qquad\qquad 
         + ~\hbox{terms including $\pd h$} ~\Bigr)~, \label{vx4}         
\eea
where ${\cal L}^{2}_{S_1}$ is derived  from the term 
$-\frac{2}{3}b_1 (\bBox \!\!\bR)\phi +\fr{1}{8}b_1 \bR^2$. 
The vertices ${\cal L}_{S_1}^{3}$ and ${\cal L}_{S_1}^{4}$ are from 
$2b_1 \phi {\bar \Delta}_4 \phi$.

\end{document}